%% file: harmonic_sa.tex
\def\confversion{0}
\def\ifconf{\ifnum\confversion=1}
\def\ifnotconf{\ifnum\confversion=0}
\documentclass[11 pt]{article}
\def\showauthornotes{1}
\def\showkeys{0}
\def\showdraftbox{1}
\usepackage{silence}
\input{macros.tex}

\usepackage[stable]{footmisc}
{}

\input{vector_macros}

\title{Hardness of Approximation for Polynomial Optimization over Sphere for Linear Programming Relaxation Hierarchy}
\author{Mrinalkanti Ghosh\thanks{%
    Indian Institute of Science, $\quad$
    \texttt{mrinalkantig@iisc.ac.in}, $\quad$
    Funding Source: Walmart Postdoc Fellowship}
  }
\begin{document}
\sloppy
\maketitle
\input{abstract}
\newpage
\tableofcontents
\newpage
\input{intro}
\input{prelims}

\input{sa_harmonic_lb}
\input{acknowledgement}
\input{ai_disclosure}
\printbibliography[
  heading=bibintoc
]
\appendix
\input{appendix}
\end{document}

%% file: macros.tex
\usepackage{amsmath,amsfonts,amssymb,amsthm,mathtools,enumerate}
\usepackage{paralist}
\usepackage{enumitem}
\usepackage{graphicx}
\usepackage{lmodern}
\usepackage{tikz}
\usetikzlibrary{arrows,calc,positioning}
\usepackage{nicefrac}
\usepackage{listings}
\usepackage{xcolor,xspace}
\usepackage[most]{tcolorbox}
\usepackage[T1]{fontenc}
\usepackage{dsfont}
\usepackage{bbm}
\usepackage{newpxmath} 

\usepackage{newpxtext} 

\usepackage{url}
\usepackage{hyperref}
\hypersetup{
    colorlinks=true,   
    pdfstartview=FitV, 
    linkcolor=blue,    
    citecolor=magenta, 
    urlcolor=green,     
}
\usepackage[
  hyperref=true,      	
  backend=biber,      
  sorting=nyt,       	
  style=alphabetic, 	
  maxalphanames=4, 
  minalphanames=3,  
  maxbibnames=999, 
  minbibnames=999
]{biblatex}

\usepackage[capitalise,nameinlink]{cleveref}

\crefname{ineq}{Ineq.}{Inequalities}
\creflabelformat{ineq}{#2{\upshape(#1)}#3}
\crefname{alg}{Algorithm}{Algorithms}
\creflabelformat{alg}{#2{\upshape(#1)}#3}
\crefname{fact}{Fact}{Facts}
\creflabelformat{fact}{#2{\upshape(#1)}#3}
\makeatletter
\AddToHook{cmd/appendix/before}{\crefalias{section}{appendix}}
\makeatother

\usepackage{thmtools}
\usepackage{thm-restate}

\usepackage{float}
\floatstyle{boxed}
\restylefloat{figure}
\usepackage{framed}

\ifnum\showkeys=1
\usepackage[color]{showkeys}
\fi

\newcommand{\calC}{\ensuremath{\mathcal{C}}\xspace}

\newcommand{\calF}{\ensuremath{\mathcal{F}}\xspace}
\newcommand{\calG}{\ensuremath{\mathcal{G}}\xspace}
\newcommand{\calH}{\ensuremath{\mathcal{H}}\xspace}

\newcommand{\calL}{\ensuremath{\mathcal{L}}\xspace}

\newcommand{\calP}{\ensuremath{\mathcal{P}}\xspace}

\newcommand{\RR}{\ensuremath{\mathbb{R}}}
\newcommand{\CC}{\ensuremath{\mathbb{C}}}

\newcommand{\EE}{\ensuremath{\mathbb{E}}}

\newcommand{\NN}{\ensuremath{\mathbb{N}}}

\renewcommand{\SS}{\ensuremath{\mathbb{S}}}

\ifnum\showauthornotes=1
\newcommand{\Authornote}[2]{{\sf\small\color{red}{[#1: #2]}}}
\newcommand{\Authorcomment}[2]{{\sf \small\color{gray}{[#1: #2]}}}
\newcommand{\Authorfnote}[2]{\footnote{\color{red}{#1: #2}}}
\else
\newcommand{\Authornote}[2]{}
\newcommand{\Authorcomment}[2]{}
\newcommand{\Authorfnote}[2]{}
\fi
\newtheorem{theorem}{Theorem}[section]

\newtheorem{lemma}[theorem]{Lemma}
\newtheorem{remark}[theorem]{Remark}
\newtheorem{proposition}[theorem]{Proposition}
\newtheorem{corollary}[theorem]{Corollary}

\newtheorem{fact}[theorem]{Fact}

\usepackage{algorithm}
\usepackage{algorithmic}

\def\FullBox{\hbox{\vrule width 6pt height 6pt depth 0pt}}

\def\qed{\ifmmode\qquad\FullBox\else{\unskip\nobreak\hfil
\penalty50\hskip1em\null\nobreak\hfil\FullBox\parfillskip=0pt\finalhyphendemerits=0\endgraf}\fi}

\def\qedsketch{\ifmmode\Box\else{\unskip\nobreak\hfil
\penalty50\hskip1em\null\nobreak\hfil$\Box$
\parfillskip=0pt\finalhyphendemerits=0\endgraf}\fi}

\ifnotconf%
  \renewenvironment{proof}{\bgroup\trivlist\item {\bf Proof:\enspace}}{\qed\endtrivlist\egroup}
\fi

\newenvironment{proofof}[1]{\begingroup\trivlist\item {\bf Proof of #1:\enspace}}{\qed\endtrivlist\endgroup}

\newcommand{\defeq}{\stackrel{\mathrm{def}}=}

\newcommand{\mper}{\,.}
\newcommand{\mcom}{\,,}

\newcommand{\abs}[1]{\ensuremath{\left\lvert #1 \right\rvert}}

\newcommand{\innerprod}[2]{\ensuremath{\left\langle #1, #2 \right\rangle}}

\newcommand{\Sch}{\mathfunction{\mathsf{S}~\!}}
\makeatletter
\def\norm#1{
  \@ifnextchar\bgroup%
   {\normalpnorm{#1}}%
   {\defaultnorm{#1}}%
}
\def\defaultnorm#1{%
    \renderNorm{#1}{}
}
\def\normalpnorm#1#2{%
   \@ifnextchar\bgroup%
   {\banaxnorm{#1}{#2}}%
   {\renderNorm{#2}{#1}}
}

\def\banaxnorm#1#2#3{%
    \renderNorm{#3}{#1\to#2 }
}

\def\renderNorm#1#2{%
    \@ifnextchar^%
    {\fixExponent{#1}{#2}}%
    {\ensuremath{\mathchoice%
        {\lVert #1 \rVert_{#2}}%
        {\lVert #1 \rVert_{#2}}%
        {\lVert #1 \rVert_{#2}}%
        {\lVert #1 \rVert_{#2}}}%
    }%
}
\def\fixExponent#1#2^#3{%
    \ensuremath{{\lVert #1 \rVert^{#3}_{#2}}}%
}

\def\enorm#1#2{%
   \@ifnextchar\bgroup%
   {\norm{L_{#1}}{L_{#2}}}%
   {\norm{L_{#1}}{#2}}%
}

\def\cnorm#1#2{%
   \@ifnextchar\bgroup%
   {\norm{\ell_{#1}}{\ell_{#2}}}%
   {\norm{\ell_{#1}}{#2}}%
}

\def\schnorm#1#2{%
   \@ifnextchar\bgroup%
   {\norm{\Sch{#1}}{\Sch{#2}}}%
   {\norm{\Sch{#1}}{#2}}%
}

\makeatother

\newcommand{\etal}{et al.\xspace}

\makeatletter

\def\HierarchyRender#1{
  \@ifnextchar\bgroup%
  {\hierarchywithlvl{#1}}
  {\nameofhierarchy{#1}}
}
\def\nameofhierarchy#1{#1\xspace}
\def\hierarchywithlvl#1#2{%
    \ensuremath{#1^{(#2)}}\!\xspace
}
\makeatother
\def\bfa{{\bf a}}
\def\bfb{{\bf b}}

\def\bfe{{\bf e}}

\def\bfv{{\bf v}}

\def\bfx{{\bf x}}
\def\bfy{{\bf y}}
\def\bfz{{\bf z}}
\def\bfX{{\bf X}}

\def\bfZ{{\bf Z}}

\def\bfphi{\ensuremath{\pmb \phi}}

\def\bfalpha{\ensuremath{\pmb \alpha}}

\newcommand{\Esymb}{\mathbb{E}}
\newcommand{\Psymb}{\mathbb{P}}

\makeatletter
\def\Pr#1{%
    \ProbabilityRender{\Psymb}{#1}%
}

\def\Ex#1{%
    \ProbabilityRender{\Esymb}{#1}%
}

\def\ProbabilityRender#1#2{
  \@ifnextchar\bgroup%
  {\renderwithdist{#1}{#2}}
   {\singlervrender{#1}{#2}}
}
\def\singlervrender#1#2{%
   \ensuremath{\mathchoice
       {{#1}\left[ #2 \right]}
       {{#1}[ #2 ]}
       {{#1}[ #2 ]}
       {{#1}[ #2 ]}
   }
}
\def\renderwithdist#1#2#3{%
   \@ifnextchar\bgroup%
   {\superfancyrender{#1}{#2}{#3}}
   {\ensuremath{\mathchoice%
      {\underset{#2}{#1}\left[ #3 \right]}
      {{#1}_{#2}[ #3 ]}
      {{#1}_{#2}[ #3 ]}
      {{#1}_{#2}[ #3 ]}
     }
   }
}
\def\superfancyrender#1#2#3#4#5{
   \ensuremath{\mathchoice
      {\underset{#1}{{#1}}\left#4 #3 \right#5}
      {{#1}_{#2}#4 #3 #5}
      {{#1}_{#2}#4 #3 #5}
      {{#1}_{#2}#4 #3 #5}
   }
}
\makeatother

\makeatletter
\def\mathfunction#1{
  \@ifnextchar\bgroup%
   {\funcWithArg{#1}}%
   {\justFunc{#1}}%
}
\def\justFunc#1{%
  \ensuremath{#1}\xspace
}
\def\funcWithArg#1#2{%
  \@ifnextchar\bgroup%
  {\funcWArgSub{#1}{#2}}%
  {\ensuremath{\mathchoice%
      {#1\!\left(#2\right)}%
      {#1\!\left(#2\right)}%
      {#1(#2)}%
      {#1(#2)}\xspace}%
  }%
}
\def\funcWArgSub#1#2#3{%
  \ensuremath{\mathchoice%
    {#1_{#2}\!\left(#3\right)}%
    {#1_{#2}\!\left(#3\right)}%
    {#1_{#2}(#3)}%
    {#1_{#2}(#3)}
  \xspace}%
}
\makeatother

\DeclareMathOperator{\divides}{\operatorname{\vert}}

\ifnum\showdraftbox=1

\else

\fi

\usetikzlibrary{decorations.markings,arrows,positioning,automata}
\tikzset{->-/.style={decoration={markings,mark=at position .5 with {\arrow{>}}},postaction={decorate}}}

%% file: vector_macros.tex
\makeatletter
\@ifundefined{ifHSAuseboldvectors}{\newif\ifHSAuseboldvectors}{}
\makeatother
\HSAuseboldvectorstrue
\providecommand{\HSAvec}[1]{\ifHSAuseboldvectors\mathbf{#1}\else#1\fi}
\providecommand{\HSAvecsymbol}[1]{\ifHSAuseboldvectors\boldsymbol{#1}\else#1\fi}

\providecommand{\bfa}{}
\renewcommand{\bfa}{\HSAvec{a}}
\providecommand{\bfb}{}
\renewcommand{\bfb}{\HSAvec{b}}

\providecommand{\bfe}{}
\renewcommand{\bfe}{\HSAvec{e}}

\providecommand{\bfv}{}
\renewcommand{\bfv}{\HSAvec{v}}

\providecommand{\bfx}{}
\renewcommand{\bfx}{\HSAvec{x}}
\providecommand{\bfy}{}
\renewcommand{\bfy}{\HSAvec{y}}
\providecommand{\bfz}{}
\renewcommand{\bfz}{\HSAvec{z}}

\providecommand{\bfX}{}
\renewcommand{\bfX}{\HSAvec{X}}

\providecommand{\bfZ}{}
\renewcommand{\bfZ}{\HSAvec{Z}}

\providecommand{\bfalpha}{}
\renewcommand{\bfalpha}{\HSAvecsymbol{\alpha}}
\providecommand{\bfbeta}{}
\renewcommand{\bfbeta}{\HSAvecsymbol{\beta}}

\providecommand{\bfxi}{}
\renewcommand{\bfxi}{\HSAvecsymbol{\xi}}
\providecommand{\bfphi}{}
\renewcommand{\bfphi}{\HSAvecsymbol{\phi}}
\providecommand{\bfzeta}{}
\renewcommand{\bfzeta}{\HSAvecsymbol{\zeta}}
\providecommand{\bfPsi}{}
\renewcommand{\bfPsi}{\HSAvecsymbol{\Psi}}


%% file: abstract.tex
\begin{abstract}
Semidefinite programming (SDP), particularly the sum-of-squares (SoS)
hierarchy, provides $n^{O(q)}$-time approximation and certification
algorithms for optimizing degree-$d$ homogeneous polynomials on the
unit sphere.  For arbitrary inputs, degree-$q$ SoS gives a powered-objective
approximation factor $O_d((n/q)^{d/2-1})$~\cite{BGGLT17}.
For random order-$d$ tensors, degree-$d$ SoS has an integrality-gap
lower bound of $2^{-O(d)}(n/d)^{d/4-1/2}$ relative to the true
spherical maximum; lower bounds also extend to growing SoS degree
in tensor PCA~\cite{BGL17,HKP17}.

We establish lower bounds for weaker linear-programming relaxation
hierarchies.  Our degree-$q$ moment LPs impose the sphere identities
and any prescribed, instance-independent family $\calG$ of valid
polynomial inequalities of degree at most $q$.  They include
Sherali--Adams-type product hierarchies.  For every even $d\ge2$,
constant $C>0$, and $q$ satisfying $2d\divides q$ and
$2d\le q\le n$, every such relaxation with at most $n^{Cq}$ tests
has instances $P$ satisfying
\[
 \frac{\operatorname{PowLP}_{q,\calG}(P)}
      {\norm{\infty}{P}}
 \ge \frac{c_{d,C}}{\sqrt{q\log n}}
       \left(\frac nq\right)^{d/2-1/2}\mper
\]
Here $\norm{\infty}{P}=\max_{\bfx\in\SS^{n-1}}\abs{P(\bfx)}$ and
$\operatorname{PowLP}_{q,\calG}(P)$ is the $d/q$ power of the
LP optimum of $\calL(P^{q/d})$.  A random spherical-harmonic
construction produces these gaps with high probability for every
fixed family of tests.  The result also applies to instance-independent
polyhedral formulations admitting degree-$q$ polynomial feasible-point
lifts and the same inequality budget.

The exponent $d/2-1/2$ improves the leading dependence on $d$
relative to the $d/4$-type exponents in prior random-tensor SoS
lower bounds.  Our LP hierarchies impose weaker positivity conditions
than semidefinite moment relaxations. It leaves open whether analogous
degree dependence can be achieved in lower bounds for SDP hierarchies.
\end{abstract}

%% file: intro.tex
\section{Introduction}
\label{sec:hsp-introduction}

Given a real homogeneous polynomial $P$ of degree $d$ on $\RR^n$,
we consider the problem of approximating
\[
 \norm{\infty}{P}\defeq
 \max_{\bfx\in\SS^{n-1}}\abs{P(\bfx)}\mper
\]
For quadratic polynomials this is the spectral norm of the associated
symmetric matrix.  Higher-degree forms give the corresponding
symmetric tensor optimization problem; already order-three tensor
spectral-norm computation is NP-hard, including for symmetric tensors~\cite{HL13}.
Approximation and certification are therefore central
questions: one seeks both a point of large value and a tractable upper
bound on the optimum.

The random ensemble is also part of an average-case lower-bound
problem.  Independent coefficients in a monomial or tensor basis need
not produce a distribution that is isotropic for the spherical $L^2$
inner product.  We instead put all randomness isotropically in the
space of degree-$d$ spherical harmonics.  We obtain a large integrality
gap for data-independent moment LPs, including Sherali--Adams-type
product relaxations.  The gap has leading degree dependence $(n/q)^{d/2-O(1)}$, as does
the general worst-case SDP approximation guarantee.  We give the
precise comparison below: the two exponents differ by an additive
$1/2$, and for a range of hierarchy degrees our result yields a
quantitative LP--SDP separation.  The lower bound is for the LP
framework, not for positive-semidefinite moment matrices.

\subsection{Polynomial optimization on the sphere: algorithms and lower bounds}
\label{subsec:hsp-literature}

\paragraph{Moment and sum-of-squares relaxations.}
The moment approach of Lasserre and the sum-of-squares approach of
Parrilo provide semidefinite hierarchies for polynomial optimization
\cite{Las01,Par00}.  On the sphere, a degree-$q$ relaxation
uses a normalized linear functional on degree-$\le q$ polynomials,
respects the sphere identities, and requires nonnegativity on every
square of degree at most $q$.  The latter condition is equivalent to
positive semidefiniteness of the moment matrix.  At fixed polynomial
degree, increasing $q$ produces a hierarchy of SDPs of size $n^{O(q)}$;
see \cite[Section 2.2]{BGL17}.  Our LPs retain finitely many
prescribed nonnegativity tests instead of requiring the full
positive-semidefinite constraint.

\paragraph{Worst-case multiplicative approximation.}
For arbitrary homogeneous degree-$d$ polynomials, He, Li, and Zhang
obtained polynomial-time approximation guarantees of order
$O_d(n^{d/2-1})$ through multilinear relaxation and polarization
\cite{HLZ10}.  A deterministic convex-geometric algorithm due to
So improves the factor to $O_d((n/\log n)^{d/2-1})$
\cite{So11}.  Here, as throughout this introduction, a factor
$A\ge1$ means that the returned value is at least the optimum divided
by $A$.

Bhattiprolu, Ghosh, Guruswami, Lee, and Tulsiani established a
running-time versus approximation tradeoff using weak decoupling and
polynomial folds \cite[Theorem 1.1]{BGGLT17}.
For $2d\divides q$ and $2d\le q\le n$, their guarantee applies to
the \emph{powered} degree-$q$ SoS bound
\[
 \operatorname{PowSoS}_q(P)
 \defeq\left(\sup_{\widetilde{\EE}}
                      \widetilde{\EE}[P^{q/d}]\right)^{d/q},
\]
\begin{samepage}
The supremum is over degree-$q$ pseudoexpectations satisfying the
sphere constraint.  Their guarantee is
\begin{equation}\label{eq:hsp-worst-case-sdp}
 \norm{\infty}{P}
 \le\operatorname{PowSoS}_q(P)
 \le C_d\left(\frac nq\right)^{d/2-1}\norm{\infty}{P}\mper
\end{equation}
\end{samepage}
Their $n^{O(q)}$-time rounding algorithm attains this factor
relative to the SDP value.  For nonnegative monomial coefficients,
they obtain $O_d((n/q)^{d/4-1/2})$ using the corresponding strengthened
relaxation.  Their sparse-polynomial guarantee is
$O_d(\sqrt{m/q})$ for $m$ monomials, in its nontrivial regime.
The distinction between powered and direct objectives matters here:
\cref{eq:hsp-worst-case-sdp} applies to the powered objective used
in this paper.

There are subsequent improvements and alternative analyses in
particular regimes.  Hsieh, Kothari, Pesenti, and Trevisan give
certifiable cubic approximation within $O(\sqrt{n/t})$ in
$2^{O(t)}\operatorname{poly}(n)$ time, including an
$O(\sqrt{n/\log n})$ polynomial-time bound, via new rounding and
compression methods \cite{HKPT24}.
Jeronimo, Wu, and Xu give a high-moment argmax approach that, among
other results, recovers the general degree-$d$ tradeoff
\cref{eq:hsp-worst-case-sdp} by a different analysis of SoS
\cite{JWX26}.  These results concern semidefinite or certifiable
rounding algorithms, rather than the fixed-test LPs studied here.

\paragraph{Convergence at large relaxation degree.}
A complementary literature studies convergence to the exact optimum.
Doherty and Wehner use a real de Finetti theorem to construct
approximate representing measures for sphere moment relaxations
\cite{DW12}.  Fang and Fawzi prove a quadratic convergence rate:
for fixed degree $d$, once the relaxation degree $q$ is at least a
suitable constant times $n$, the direct-objective SoS error, divided
by the range $P_{\max}-P_{\min}$, is $O_d(n^2/q^2)$
\cite[Theorem 1]{FF21}.  This range-normalized statement is
different from the multiplicative powered-objective bound above.
Lovitz and Johnston construct a convergent spectral hierarchy whose
individual levels require an eigenvalue computation instead of a
full SDP \cite{LJ26}.  The coarse-approximation regime
$q\le n$ considered here is distinct from asymptotic convergence
with $q$ large relative to the dimension.

\paragraph{Random inputs and semidefinite lower bounds.}
For fixed-degree random tensors with independent standard Gaussian
or Rademacher entries, the true spherical maximum is
$O_d(\sqrt n)$ with high probability, whereas constant-degree
certificates can be much larger; see \cite{BGL17}.
Hopkins, Shi, and Steurer established the characteristic
$n^{3/4}$ signal scale for degree-four SoS in cubic tensor PCA,
together with a corresponding lower bound up to logarithmic factors
\cite{HSS15}.  Bhattiprolu, Guruswami, and Lee proved that
degree-$q$ SoS certifies random degree-$d$ tensor maxima within
\begin{equation}\label{eq:hsp-random-tensor-sdp}
 \widetilde O_d\left(\left(\frac nq\right)^{d/4-1/2}\right)
\end{equation}
and established matching basic-degree gaps up to degree-dependent
factors \cite[Theorems 1.1--1.2]{BGL17}.
Hopkins et al.\ obtained higher-degree lower bounds for tensor PCA
using moment-matching constructions; these exhibit the same
$d/4$-type dependence, with losses in the hierarchy parameter
\cite[Section 6]{HKP17}.  Their Gaussian transfer in Appendix C
is an expectation statement, so we do not identify it with a
high-probability Gaussian lower bound.

The present result differs in both the random input and the relaxation
class.  We use isotropic spherical harmonics rather than independent
tensor entries and prove a $d/2$-type exponent for prescribed moment
LPs.  The product-and-linearization construction of Sherali and Adams
\cite{SA90} supplies natural examples of such LPs on the sphere.
The finite-family theorem also allows other valid, data-independent
polynomial tests.  It is not a sum-of-squares lower bound.

\subsection{The random model and the main theorem}
\label{subsec:hsp-intro-result}

Let $\sigma$ be uniform probability measure on $\SS^{n-1}$, and let
$\calH_{n,d}$ be the real vector space of homogeneous harmonic
polynomials of degree $d$, restricted to the sphere.  Its dimension is
\[
 h=h_{n,d}=\binom{n+d-1}{d}-\binom{n+d-3}{d-2}\mper
\]
Choose any $L^2(\sigma)$-orthonormal basis $Y_1,\ldots,Y_h$ and sample
\[
 P_{\bfa}(\bfx)=\sum_{i=1}^h a_iY_i(\bfx),
 \qquad \bfa\sim\operatorname{Unif}(\SS^{h-1})\mper
\]
This distribution is independent of the choice of orthonormal basis.
It is equivalently obtained by taking a standard Gaussian in
$\calH_{n,d}$ and dividing by its $L^2(\sigma)$ norm.  In particular,
$\norm{L^2(\sigma)}{P_{\bfa}}=1$ for every sample.  The harmonic
addition formula implies that $P_{\bfa}(\bfx)$ has variance one at
each fixed $\bfx$ on the sphere; see \cite[Sections 1--2]{DX13}
and \cref{subsec:hsp-harmonics}.  A standard norming-net argument gives
$\norm{\infty}{P_{\bfa}}\le C_d\sqrt n$ with probability at least
$1-e^{-n}$; we record the proof in \cref{lem:hsp-true-optimum}.

The relaxation uses moments through degree $q$, where $2d\divides q$.
Writing $r=q/d\in2\NN$, its objective is $\calL(P_{\bfa}^r)$, and
we take the $r$th root only after maximizing.  This is the
powered-objective convention used in
\cite[Theorem 1.1 and Section 4.2]{BGGLT17}.  Given a family
$\calG$ of polynomials of degree at most $q$ that are nonnegative on
the sphere, write
\[
 \operatorname{PowLP}_{q,\calG}(P_{\bfa})
 =\left(\sup_{\calL\in\calF_{q,\calG}}
                   \calL(P_{\bfa}^r)\right)^{1/r}\mper
\]
The region $\calF_{q,\calG}$ imposes normalization, the linear sphere
equalities, and $\calL(g)\ge0$ for $g\in\calG$; its formal definition
is given in subsection~\ref{subsec:hsp-relaxation}.
Both the tests and the feasible region are fixed independently of the
observed coefficient vector $\bfa$.

\begin{theorem}[Main theorem: Sherali--Adams-type gaps]
\label{thm:hsp-gap}
\label{cor:hsp-gap} 
Fix an even integer $d\ge2$ and constants $C,B>0$.  There is a
constant $c_{d,C,B}>0$ with the following property.  Let $n\ge2d$,
let $q$ satisfy $2d\divides q$ and $2d\le q\le n$, and fix a family
$\calG\subseteq\RR[\bfx]_{\le q}$ of at most $n^{Cq}$ polynomials
nonnegative on $\SS^{n-1}$, independently of $\bfa$.  Then, with
probability at least $1-n^{-B}-e^{-n}$ over the random harmonic
polynomial $P_{\bfa}$ defined above,
\begin{equation}\label{eq:hsp-gap}
 \frac{\operatorname{PowLP}_{q,\calG}(P_{\bfa})}
      {\norm{\infty}{P_{\bfa}}}
 \ge\frac{c_{d,C,B}}{\sqrt{q\log n}}
             \left(\frac nq\right)^{d/2-1/2}\mper
\end{equation}
The probability statement is for each prescribed family $\calG$;
it is not a simultaneous assertion over all such families.
\end{theorem}

For fixed $q$, the gap is
$\Omega_{d,q,C,B}(n^{d/2-1/2}/\sqrt{\log n})$.
At larger $q$, the bound retains the power $d/2-1/2$ of $n/q$, with
an additional factor $1/\sqrt{q\log n}$.  We do not claim that the
dependence on $q$ is optimal.  The theorem allows arbitrary prescribed
valid degree-$q$ inequalities, not just square tests or a particular
choice of SA generators.  An empty test family can be augmented by
the redundant test $1\ge0$.

Products of a polynomial-size collection of valid generators, through
total degree $q$, give $n^{O(q)}$ tests and are covered by the theorem.
For example, one can use products of the box inequalities $1\pm x_i$
and impose the sphere equalities in addition.  This produces a bounded
moment LP.  More generally, the result applies to data-independent LP
formulations admitting degree-$q$ polynomial feasible-point lifts; see
\cref{subsec:hsp-lifts}.  The degree restriction on the lift is part of
the theorem's scope.  We do not assert a size lower bound for arbitrary
extended formulations with unrestricted feasible-point lifts.

\subsection{Comparison with the SDP approximation guarantee}
\label{subsec:hsp-comparison}

\paragraph{Near-matching degree dependence, but different hierarchies.}
At the same $n^{O(q)}$ size scale, the worst-case SDP guarantee
\cref{eq:hsp-worst-case-sdp} has factor
$(n/q)^{d/2-1}$, whereas the LP gap in \cref{thm:hsp-gap} is
\[
 \frac{c_{d,C,B}}{\sqrt{q\log n}}
       \left(\frac nq\right)^{d/2-1/2}\mper
\]
Thus both have leading dependence $(n/q)^{d/2-O(1)}$ on the
polynomial degree, in contrast to the $d/4$-type exponents in the
random-tensor SoS bounds.  We obtain this dependence while enforcing
only prescribed LP inequalities, rather than positive semidefiniteness.
The exponents $d/2-1$ and $d/2-1/2$ are nevertheless different:
for fixed $d$, their ratio contributes a genuine factor $\sqrt{n/q}$,
not a degree-dependent constant or a polylogarithmic loss.
Accordingly, the comparison is near-matching in its leading
\emph{degree dependence}, not a claim of matching upper and lower
bounds for the same hierarchy or of an optimal rate in $q$.

\paragraph{A quantitative LP--SDP separation.}
Apply the deterministic guarantee \cref{eq:hsp-worst-case-sdp}
to the same random harmonic polynomial as in \cref{thm:hsp-gap}.
On the event in that theorem,
\begin{equation}\label{eq:hsp-lp-sdp-separation}
 \frac{\operatorname{PowLP}_{q,\calG}(P_{\bfa})}
      {\operatorname{PowSoS}_q(P_{\bfa})}
 \ge c'_{d,C,B}\frac{\sqrt n}{q\sqrt{\log n}}\mper
\end{equation}
Indeed,
\[
 \frac{(n/q)^{d/2-1/2}}{\sqrt{q\log n}}
 =\left(\frac nq\right)^{d/2-1}
        \frac{\sqrt n}{q\sqrt{\log n}}\mper
\]
For fixed $q$, the ratio in \cref{eq:hsp-lp-sdp-separation} grows
at least as $\sqrt{n/\log n}$ up to fixed parameters.  More generally,
it diverges whenever $q=o(\sqrt{n/\log n})$.  This quantifies a loss incurred by the prescribed LP positivity
constraints rather than an obstruction to SDP optimization.

For literal containment of feasible regions, one can add the same
tests $\calG$ to the SoS relaxation.  This preserves every evaluation
functional, only decreases the SDP optimum, and makes it a
strengthening of the LP.  The same separation therefore holds for
this strengthened SDP.  This qualification is relevant for arbitrary
valid tests that need not themselves be sums of squares.

These comparisons do not show that the harmonic ensemble is harder
for all algorithms than the tensor ensemble.  Already for $d=2$,
the exact optimization problem is a matrix eigenvalue problem, while
our theorem still gives a large gap for prescribed LP tests.  The
obstruction established here is the restricted positivity enforced
by these formulations.

\subsection{Why random spherical harmonics?}
\label{subsec:hsp-ensemble-motivation}

A further, heuristic motivation for studying pure random spherical harmonics
is the possibility of a complicated optimization landscape with many competing
local maxima.  For a Gaussian spherical harmonic $P$ of degree $d$ on
$\SS^{n-1}$, let $Z(P)=\{\bfx\in\SS^{n-1}:P(\bfx)=0\}$ and let
$b_0(Z(P))$ count its connected components.  When $n$ is fixed and
$d\to\infty$, the expected number of nodal components is
$\Ex{}{b_0(Z(P))}=\Theta_n(d^{n-1})$; this has the maximal order in
$d$ allowed by classical algebraic-topological bounds
\cite{NS09,LL15}.  Nodal components are not local maxima, but they
reflect an oscillatory landscape.  More directly, the expected number
of local extrema for pure Gaussian spherical harmonics also has order
$d^{n-1}$ \cite[Section~8, Example~4]{FLL15}.  By symmetry under
$P\mapsto-P$, the expected numbers of local maxima and minima are equal.
The resulting abundance of competing peaks suggests that this family is
a natural candidate for difficult global maximization, although it does
not in itself establish computational hardness.

This motivation is tangential to our lower-bound proof, and the
asymptotic regimes are different.  The preceding topological and
critical-point results hold with the ambient dimension $n$ fixed and
the degree $d$ increasing; we instead fix $d$ and let $n\to\infty$.
Consequently, those results neither imply nor predict the LP gap proved
here.  They only provide independent intuition for choosing the
pure-harmonic ensemble.  Since normalization by a positive scalar does
not alter local maxima or nodal sets, this intuition applies equally
to our coefficient-sphere normalization.

To make the comparison with monomial randomness precise, consider the
rotation-invariant Gaussian tensor ensemble
\begin{equation}\label{eq:hsp-kostlan-model}
 G(\bfx)=\sum_{\abs{\bfbeta}=d}
       \sqrt{\binom d{\bfbeta}}\,g_{\bfbeta}\bfx^{\bfbeta},
 \qquad g_{\bfbeta}\sim N(0,1)\ \text{independently},
 \qquad \binom d{\bfbeta}=\frac{d!}{\prod_i\beta_i!}\mper
\end{equation}
Grouping the ordered entries of an independent Gaussian order-$d$
tensor gives exactly this law.  Its covariance is
\[
 \Ex{}{G(\bfx)G(\bfy)}=\innerprod{\bfx}{\bfy}^{d}\mper
\]
This is the Kostlan--Shub--Smale ensemble, in the normalization with
pointwise variance one on the sphere.  Its harmonic decomposition is
classical; see \cite[Proposition 3.1]{Gic17} and
\cite[Section 8, Example 1]{FLL15}.  Equal variances on the
\emph{distinct unweighted monomials} define a different, generally
non-rotation-invariant law.  The degree-only spectral weights below
refer specifically to \cref{eq:hsp-kostlan-model}.

Let $\calC_{\mathrm K}$ be the spherical covariance operator of $G$.
On the harmonic space $\calH_{n,j}$, for $0\le j\le d$ and
$j\equiv d\pmod2$, its eigenvalue is
\begin{equation}\label{eq:hsp-kostlan-eigenvalues}
 \lambda_{n,d,j}
 =\frac{d!}
 {2^{(d-j)/2}((d-j)/2)!\,\mathsf R_{n,(d+j)/2}},
 \qquad \mathsf R_{n,k}=\prod_{t=0}^{k-1}(n+2t)\mper
\end{equation}
Equivalently, in an orthonormal spherical-harmonic basis, each
coefficient at harmonic degree $j$ has variance $\lambda_{n,d,j}$.
Formula \cref{eq:hsp-kostlan-eigenvalues} is the covariance
normalization of the rescaling factors in
\cite[Proposition 3.1, equation (3.11)]{Gic17}.
A direct spherical-Wick verification is given in
\cref{app:hsp-kostlan-spectrum}.
For consecutive admissible harmonic degrees,
\begin{equation}\label{eq:hsp-kostlan-ratio}
 \frac{\lambda_{n,d,j-2}}{\lambda_{n,d,j}}
 =\frac{n+d+j-2}{d-j+2}>1
 \qquad(2\le j\le d)\mper
\end{equation}
Thus the monomial/tensor model is \emph{bottom-heavy per harmonic
coefficient}: a unit $L^2$ direction at a lower harmonic degree
receives more variance than a unit direction at a higher degree.

Per-coefficient variance must be distinguished from the total mass
of a harmonic level.  That mass is
$h_{n,j}\lambda_{n,d,j}$.  In the regime of this paper, with fixed
$d$ and growing $n$, the top-level mass satisfies
\begin{equation}\label{eq:hsp-kostlan-top-mass}
 h_{n,d}\lambda_{n,d,d}
 =\frac{h_{n,d}d!}{\mathsf R_{n,d}}
 =1-O_d(n^{-1})\mper
\end{equation}
Thus the total mass is not concentrated at the bottom in this regime.
In the different regime of fixed $n$ and growing $d$, Kostlan mass
is instead concentrated at harmonic degrees on the order of
$\sqrt d$; a precise scaling limit and approximation by lower-degree
polynomials are established in
\cite[Theorems 4.2 and 5.3]{Gic17}.

The spherical $L^2$-isotropic, or real Fubini--Study, ensemble assigns
equal variance to orthonormal harmonic coefficients across the degrees
$d,d-2,\ldots$ \cite[Section 8, Example 2]{FLL15}.
Our monochromatic model puts all variance uniformly on its largest
harmonic summand, $\calH_{n,d}$, whose dimension is
$\Theta_d(n^d)$.  It removes the preferential variance assigned to
lower-degree directions.  This distinction can also be quantified
through the covariance participation ratio
\[
 D_{\mathrm{eff}}(\calC)
 \defeq\frac{(\operatorname{tr}\calC)^2}
                  {\operatorname{tr}(\calC^2)}\mper
\]
Both covariance operators have trace one.  For our ensemble,
$\calC_{\mathrm{harm}}=\Pi_d/h$, so
$D_{\mathrm{eff}}(\calC_{\mathrm{harm}})=h$.
For the tensor ensemble, spherical Wick gives
\begin{equation}\label{eq:hsp-effective-dimension}
 \begin{aligned}
 \operatorname{tr}(\calC_{\mathrm K}^2)
 &=\iint\innerprod{\bfx}{\bfy}^{2d}
                    \,\mathrm d\sigma(\bfx)\mathrm d\sigma(\bfy)
   =\frac{(2d-1)!!}{\mathsf R_{n,d}},\\
 \frac{D_{\mathrm{eff}}(\calC_{\mathrm{harm}})}
      {D_{\mathrm{eff}}(\calC_{\mathrm K})}
 &\longrightarrow\frac{(2d-1)!!}{d!}
 \qquad(n\to\infty,\ d\text{ fixed})\mper
 \end{aligned}
\end{equation}
This describes a more uniform spread across high-degree directions,
not an increase in the algebraic support dimension: the full
Kostlan ensemble has support of dimension $\binom{n+d-1}{d}$.
Indeed, projecting a Kostlan sample onto $\calH_{n,d}$ and normalizing
its spherical $L^2$ norm gives exactly our model.
For fixed $d$, the effective-dimension improvement in
\cref{eq:hsp-effective-dimension} is a constant depending on $d$,
not a new power of $n$.

These spectral considerations motivate isotropic high-harmonic inputs
as candidates for strong average-case lower bounds.  They are not,
by themselves, a computational-hardness argument or a proof that the
harmonic ensemble is harder than the tensor ensemble.  The lower bound
in \cref{thm:hsp-gap} follows from the explicit feasible functional
constructed below.

\subsection{Proof overview}
\label{subsec:hsp-intro-overview}

Set $H=h_{n,q}$ and let
$F_{\bfa}=\Pi_q(P_{\bfa}^{r})$ be the top solution-harmonic component
of the powered objective.  For a parameter $\alpha\ge0$ we use
\[
 \calL_{\bfa,\alpha}(f)
 =\int f(\bfx)\bigl(1+\alpha F_{\bfa}(\bfx)\bigr)
                  \,\mathrm d\sigma(\bfx)\mper
\]
This construction is motivated by spherical moment matching, in the
spirit of pseudo-calibration \cite[Section 3]{HKP17}.
For the LP argument we need only the displayed signed functional,
not a full calibrated distribution.  Its constant term is normalized,
and the sphere equalities hold for every $\alpha$.  Since $r$ is even,
\[
 \calL_{\bfa,\alpha}(P_{\bfa}^r)
 =\int P_{\bfa}^r\,\mathrm d\sigma
      +\alpha\norm{L^2(\sigma)}{F_{\bfa}}^2
 \ge\alpha\norm{L^2(\sigma)}{F_{\bfa}}^2\mper
\]
We therefore maximize the useful perturbation scale $\alpha$ subject
to the fixed tests.

Two estimates suffice.  First, spherical Wick and harmonic
orthogonality give an exact covariance identity for $F_{\bfa}$.
For any fixed test $g$, the random quantity
$Z_g(\bfa)=\int gF_{\bfa}\,\mathrm d\sigma$ is a homogeneous
polynomial of degree $r$ in the data vector, and
\[
 \Ex{}{Z_g(\bfa)^2}
 \le\frac{r!}{H}\norm{L^2(\sigma)}{\Pi_qg}^2\mper
\]
For $g\ge0$ of degree at most $q$, standard hypercontractivity gives
$\norm{L^2(\sigma)}{g}\le3^q\int g\,\mathrm d\sigma$.
Hypercontractivity on the coefficient sphere then controls higher
moments of $Z_g$.  This step uses only degree-$r$ homogeneity,
not harmonicity in the data variables.  Taking a moment order of size
$\max\{2,\log(M/\delta)\}$ and applying a union bound determines
an admissible value of $\alpha$ simultaneously for all $M$ tests.

Second, we prove the deterministic estimate
\[
 \norm{L^2(\sigma)}{\Pi_q(P^r)}^2\ge r!\,c_d^{\,r}
 \qquad\bigl(P\in\calH_{n,d},\ \norm{L^2(\sigma)}{P}=1,\ q=dr\le n\bigr),
\]
where $c_d>0$ depends only on $d$.
The proof restricts $P$ to a complex isotropic subspace, on which
all trace terms vanish.  Rotation averaging supplies a restriction
with sufficient energy, and the standard holomorphic Gaussian moment
estimate from polarization and decoupling supplies the factor $r!$;
see \cref{lem:hsp-complex-moments,lem:hsp-top-norm}.
Combining these estimates gives the finite-family main lemma
\cref{lem:hsp-main}, and the standard true-supremum bound yields
\cref{thm:hsp-gap}.  No power inequality for signed functionals and
no positive-semidefiniteness assertion are used.

\paragraph{Organization.}
\Cref{sec:hsp-preliminaries} defines the relaxation and records the
standard harmonic, moment, and hypercontractive tools.
\Cref{sec:hsp-contribution} defines the parameterized witness, proves
the main lemma and theorem, and treats SA-type product families and
polynomial feasible-point lifts.
\Cref{app:hsp-spherical-tensors} contains the self-contained
spherical-Wick and tensor-inner-product proofs, and
\cref{app:hsp-kostlan-spectrum} verifies the harmonic-weight comparison.

%% file: prelims.tex
\section{Preliminaries for powered polynomial optimization}
\label{sec:hsp-preliminaries}

\paragraph{Vector notation.}
We use boldface symbols for vectors, such as $\bfa$, $\bfx$, and
$\bfZ$, and unbolded symbols for their scalar coordinates.  For example,
$\bfa=(a_1,\ldots,a_h)$ and $\bfx=(x_1,\ldots,x_n)$.
Scalar parameters, including $\alpha$, are also unbolded; in particular,
$\bfa$ denotes the data vector, whereas $\alpha$ is a scalar.
Vector-valued maps and multi-indices follow the same boldface convention.
Matrices and tensors retain their individually specified notation.

We collect the standard harmonic-analytic, probabilistic, and tensor
identities needed below.  The lower-bound construction and its
application-specific estimates are separated into
\cref{sec:hsp-contribution}.  Throughout, $n\ge3$, $d\ge2$ is even,
$q\ge2d$, $2d\divides q$, and
\[
 r\defeq q/d\in2\NN\mper
\]
The asymptotic results assume $q\le n$ and fixed $d$.  All measures on
spheres are probability measures; all logarithms are natural.  The
spherical--tensor inner-product identity and its orbit-averaging
interpretation are recorded here; the auxiliary identities and
self-contained proofs are collected in
\cref{app:hsp-spherical-tensors}.

\subsection{The powered-objective moment relaxation}
\label{subsec:hsp-relaxation}
\label{subsec:hsp-powered-objective}
\label{subsec:hsp-main}

For a degree-$d$ homogeneous polynomial $P$, the identity
\begin{equation}\label{eq:hsp-true-powered}
 \max_{\bfx\in\SS^{n-1}}P(\bfx)^r
 =\left(\max_{\bfx\in\SS^{n-1}}\abs{P(\bfx)}\right)^r
\end{equation}
uses that $r$ is even.  A degree-$q$ relaxation of this problem
optimizes the linear functional $\calL(P^r)$ and takes the $r$th root
\emph{after} optimization.  This is the powered-objective convention
used, for example, in \cite[Theorem 1.1 and Section 4.2]{BGGLT17}.
For an LP in moment variables, $\calL(P^r)$ remains a linear objective;
its coefficients are degree-$r$ polynomials of the observed coefficients
of $P$.

This formulation is different from optimizing $\calL(P)$ over the
same degree-$q$ moment region: an arbitrary signed functional need not
satisfy $\calL(P^r)\ge\calL(P)^r$.  We work throughout with the powered
objective itself.

Fix, independently of the observed vector $\bfa$, a finite family
$\calG=\{g_1,\ldots,g_M\}\subseteq\RR[\bfx]_{\le q}$ with $M\ge1$ and
$g_j\ge0$ on $\SS^{n-1}$.  Let $\calF_{q,\calG}$ be the region of
linear functionals $\calL:\RR[\bfx]_{\le q}\to\RR$ satisfying
\begin{equation}\label{eq:hsp-feasible-region}
 \begin{aligned}
 \calL(1)&=1,\\
 \calL((\norm{\bfx}^2-1)f)&=0
                  &&(\deg f\le q-2),\\
 \calL(g_j)&\ge0   &&(j\in[M])\mper
 \end{aligned}
\end{equation}
Define the powered-objective LP value by
\begin{equation}\label{eq:hsp-powered-lp}
 \operatorname{PowLP}_{q,\calG}(P)
 \defeq\left(\sup_{\calL\in\calF_{q,\calG}}
                    \calL(P^r)\right)^{1/r}\mper
\end{equation}
The inner problem is a linear program.  Every evaluation functional
at a point on the sphere is feasible.  Thus its supremum is positive
for nonzero $P$, possibly $+\infty$, and
$\operatorname{PowLP}_{q,\calG}(P)\ge\norm{\infty}{P}$.
Bounded SA-type product examples and polynomial feasible-point lifts are
specified in \cref{subsec:hsp-lifts}.

\paragraph{The random harmonic model.}
Let $\sigma$ be uniform probability measure on $\SS^{n-1}$.
For an integer $j\ge0$, let $\calH_{n,j}$ denote the space of real
homogeneous harmonic polynomials of degree $j$, identified with their
restrictions to the sphere, as in \cite[Definition~1.1]{DX13}.
Its dimension is, by \cite[Corollary~1.4]{DX13},
\begin{equation}\label{eq:hsp-dimension}
 h_{n,j}=\binom{n+j-1}{j}-\binom{n+j-3}{j-2}\mper
\end{equation}
The second term is zero for $j<2$.  Set
\[
 h\defeq h_{n,d}\mcom\qquad H\defeq h_{n,q}\mper
\]
Fix an orthonormal basis $Y_1,\ldots,Y_h$ of $\calH_{n,d}$ in
$L^2(\sigma)$, and write
\begin{equation}\label{eq:hsp-model}
 \bfPsi_d(\bfx)=(Y_1(\bfx),\ldots,Y_h(\bfx))\mcom\qquad
 P_{\bfa}(\bfx)=\innerprod{\bfa}{\bfPsi_d(\bfx)}\mcom\qquad
 \bfa\sim\nu\defeq\operatorname{Unif}(\SS^{h-1})\mper
\end{equation}
Thus $P_{\bfa}$ is a Gaussian harmonic polynomial divided by its spherical
$L^2$ norm.  The null law is exactly the coefficient-sphere law, not a
product Gaussian approximation.

\subsection{Harmonic projections and reproducing kernels}
\label{subsec:hsp-harmonics}

We use standard facts about spherical harmonics and Gegenbauer
polynomials with the normalization of Dai--Xu~\cite{DX13}.  In
particular, orthogonality of different harmonic degrees is
\cite[Theorem~1.2]{DX13}, the Fischer decomposition of homogeneous
polynomials is \cite[Theorem~1.3]{DX13}, and the dimension formula
below is \cite[Corollary~1.4]{DX13}.  We record the consequences
needed later in our notation.

The standard harmonic decomposition gives
\begin{equation}\label{eq:hsp-decomposition}
 \RR[\bfx]_{\le k}\big|_{\SS^{n-1}}
 =\bigoplus_{j=0}^k\calH_{n,j}\mper
\end{equation}
For a homogeneous degree-$k$ polynomial $G$, there is a unique
homogeneous harmonic polynomial $\Pi_kG$ such that
\begin{equation}\label{eq:hsp-fischer}
 G(\bfx)=\Pi_kG(\bfx)+(x_1^2+\cdots+x_n^2)R(\bfx),
 \qquad \deg R=k-2\mper
\end{equation}
This is the top-degree term in the Fischer decomposition.  On the
sphere, $\Pi_k$ is orthogonal projection onto $\calH_{n,k}$; in
particular, it annihilates every polynomial restriction of degree
strictly less than $k$.  These statements also follow from
\cite[Theorem~1.3 and Corollary~1.5]{DX13}.

In ambient dimension $m\ge3$, put
\begin{equation}\label{eq:hsp-gegenbauer}
 Q_j^{(m)}(t)=\frac{C_j^{(m-2)/2}(t)}{C_j^{(m-2)/2}(1)},
 \qquad
 \ell_{m,j}\defeq[t^j]Q_j^{(m)}(t)
       =2^j\frac{((m-2)/2)_j}{(m-2)_j}\mper
\end{equation}
Here $(b)_j=b(b+1)\cdots(b+j-1)$, with $(b)_0=1$.
The normalization and leading coefficient in
\cref{eq:hsp-gegenbauer} follow from the standard Gegenbauer formula
\cite[(2.6)]{DX13} together with the value at $1$ implicit in
\cite[Theorem~2.6 and Corollary~2.7]{DX13}.  With this
normalization, the addition formula
\cite[Theorem~2.6, equation~(2.8)]{DX13} says that the reproducing
kernel for $\calH_{n,j}$ is
\begin{equation}\label{eq:hsp-kernel}
 K_j(\bfx,\bfy)=h_{n,j}Q_j^{(n)}(\innerprod{\bfx}{\bfy})\mper
\end{equation}
In particular, the diagonal identity
$K_j(\bfx,\bfx)=h_{n,j}$ is exactly
\cite[Corollary~2.7]{DX13}.  Consequently,
\begin{equation}\label{eq:hsp-basic-normalizations}
 \norm{\bfPsi_d(\bfx)}^2=h,\qquad
 \int\bfPsi_d\bfPsi_d^\top\,\mathrm d\sigma=I_h,\qquad
 \int P_{\bfa}\,\mathrm d\sigma=0,\qquad
 \int P_{\bfa}^2\,\mathrm d\sigma=1\mper
\end{equation}
The first and last identities hold pointwise for $\bfx\in\SS^{n-1}$
and $\bfa\in\SS^{h-1}$, respectively.  Homogeneous extension gives
\begin{equation}\label{eq:hsp-sum-squares}
 \sum_{i=1}^hY_i(\bfx)^2=h(x_1^2+\cdots+x_n^2)^d
 \qquad(\bfx\in\RR^n)\mper
\end{equation}

We will use the radial-moment notation
\begin{equation}\label{eq:hsp-radial}
 \mathsf R_{m,j}\defeq\prod_{t=0}^{j-1}(m+2t),
 \qquad \mathsf R_{m,0}=1\mper
\end{equation}
The dimension and leading-coefficient formulas imply
\begin{equation}\label{eq:hsp-leading-identity}
 h_{m,j}\ell_{m,j}=\frac{\mathsf R_{m,j}}{j!}\mper
\end{equation}
Also, $\ell_{n,0}=1$ and
\[
 \frac{\ell_{n,j+1}}{\ell_{n,j}}
   =\frac{n-2+2j}{n-2+j}
\]
is increasing in $j$.  Hence $\ell_{n,u+v}\ge\ell_{n,u}\ell_{n,v}$,
and in particular
\begin{equation}\label{eq:hsp-beta-bound}
 0<\frac{\ell_{n,d}^r}{\ell_{n,q}}\le1\mper
\end{equation}
For $2\le q\le n$, the dimension identity also gives
\begin{equation}\label{eq:hsp-large-dimension}
 H=\frac{n+2q-2}{n+q-2}\binom{n+q-2}{q}
 \ge\left(\frac nq\right)^q\mper
\end{equation}
Indeed, $\binom{N}{q}\ge(N/q)^q$ and $n+q-2\ge n$.

\subsection{Tensor identities, irreducibility, and orbit averaging}
\label{subsec:hsp-tensors}

All tensor norms use the Frobenius inner product, extended as a
Hermitian inner product for complex tensors.  We use the convention
that this inner product is linear in its first argument:
\[
 \innerprod{S}{T}_{\mathrm F}
 \defeq\sum_{i_1,\ldots,i_k=1}^n
 S_{i_1,\ldots,i_k}\overline{T_{i_1,\ldots,i_k}}\mper
\]
A homogeneous polynomial $G$ of degree $k$ has a unique symmetric
coefficient tensor $T_G$ characterized by
\[
 G(\bfx)=\sum_{i_1,\ldots,i_k=1}^n(T_G)_{i_1,\ldots,i_k}
             x_{i_1}\cdots x_{i_k},
 \qquad
 (T_G)_{i_1,\ldots,i_k}
 =\frac{1}{k!}\partial_{i_1}\cdots\partial_{i_k}G\mper
\]
For $k=0$ the tensor is a scalar.  For $k\ge2$, define its trace by
\[
 (\operatorname{Tr}_{2}T_G)_{i_3,\ldots,i_k}
 \defeq\sum_{j=1}^n(T_G)_{j,j,i_3,\ldots,i_k}\mper
\]
Symmetry makes the choice of the two traced slots immaterial.
Tensors of order zero or one are trace-free by convention.

\begin{lemma}[Spherical and tensor inner products]
\label{lem:hsp-tensor-sphere}
Let $n\ge2$, let $k\ge0$ be an integer, and let $\sigma$ be uniform
\emph{probability} measure on $\SS^{n-1}$.  Put
$\mathsf R_{n,k}=\prod_{j=0}^{k-1}(n+2j)$, with
$\mathsf R_{n,0}=1$.  If $G$ is a real or complex homogeneous harmonic
polynomial of degree $k$, then
\begin{equation}\label{eq:hsp-tensor-sphere}
 \norm{L^2(\sigma)}{G}^2
 =\frac{k!}{\mathsf R_{n,k}}\norm{\mathrm F}{T_G}^2\mper
\end{equation}
More generally, if $A$ is any homogeneous polynomial of degree $k$
and $B$ is a homogeneous harmonic polynomial of degree $k$, both
possibly complex, then
\begin{equation}\label{eq:hsp-tensor-sphere-polarized}
 \int A(\bfx)\overline{B(\bfx)}\,\mathrm d\sigma(\bfx)
 =\frac{k!}{\mathsf R_{n,k}}
       \innerprod{T_A}{T_B}_{\mathrm F}\mper
\end{equation}
In particular, only one of the two polynomials needs to be harmonic
in the mixed identity.
\end{lemma}

The mixed identity is the tensor form of the Fischer--spherical
inner-product identity in \cite[Theorem 1.8(3)]{DX13}.
A self-contained proof using the spherical Wick formula is given in
\cref{app:hsp-tensor-sphere-proof}.  The supporting identities and
their proofs are collected in \cref{app:hsp-spherical-tensors}.

\paragraph{Irreducibility and orbit averaging.}
The representation-theoretic input is the following classical theorem.
We state its standard complex form in our notation, with ambient
dimension $n$ and harmonic degree $k$.  The complex formulation is justified by the same reproducing-kernel
argument as their proof; we record this point in
\cref{app:hsp-orbit-proof}.  It does not follow merely from real
irreducibility for an arbitrary representation.

\begin{theorem}[Dai--Xu, {\cite[Theorem 7.2]{DX13}}]
\label{thm:hsp-harmonic-irreducibility}
Let $n\ge3$ and $k\ge0$.  On the complexification
$\calH_{n,k}^{\CC}\defeq\calH_{n,k}\otimes_{\RR}\CC$, the action
\[
 (\rho_{n,k}(O)G)(\bfx)\defeq G(O^{-1}\bfx),
 \qquad O\in\operatorname{SO}(n),
\]
is an irreducible unitary representation for the spherical
$L^2$ inner product.  In particular, its only invariant complex
linear subspaces are $\{0\}$ and $\calH_{n,k}^{\CC}$.
\end{theorem}

The same action is irreducible for $\operatorname{O}(n)$: an
$\operatorname{O}(n)$-invariant subspace is also
$\operatorname{SO}(n)$-invariant.  Under the correspondence
$G\leftrightarrow T_G$, the space $\calH_{n,k}^{\CC}$ is identified
with the space $E_{n,k}$ of symmetric trace-free complex tensors of
order $k$, and the action becomes
$U_{n,k}(O)\defeq O^{\otimes k}|_{E_{n,k}}$.
The spherical and Frobenius inner products differ by the positive
scalar in \cref{eq:hsp-tensor-sphere}; thus this identification also
transfers unitarity and irreducibility to the harmonic tensor space.
The following consequence, rather than irreducibility by itself, is
the averaging statement used in the lower-bound proof.

\begin{fact}[Orbit averaging on a harmonic degree]
\label{fact:hsp-harmonic-averaging}
\label{fact:hsp-irreducibility} 
Let $n\ge3$, $k\ge0$, and $h_{n,k}=\dim_{\CC}E_{n,k}$.
Let $\mu$ be Haar probability measure on either
$\operatorname{SO}(n)$ or $\operatorname{O}(n)$, denoted below by $K$.
For every Hermitian operator $A:E_{n,k}\to E_{n,k}$,
\begin{equation}\label{eq:hsp-harmonic-operator-average}
 \int_K U_{n,k}(O)A U_{n,k}(O)^*\,\mathrm d\mu(O)
 =\frac{\operatorname{tr}A}{h_{n,k}}I_{E_{n,k}}\mper
\end{equation}
In particular, if $W\subseteq E_{n,k}$ is a complex subspace of
dimension $b$, $A_W$ is its Hermitian orthogonal projection, and
$T\in E_{n,k}$, then
\begin{equation}\label{eq:hsp-harmonic-projector-average}
 \int_K A_{U_{n,k}(O)W}\,\mathrm d\mu(O)
 =\frac{b}{h_{n,k}}I_{E_{n,k}},
 \qquad
 \Ex{O\sim\mu}{\norm{\mathrm F}{A_{U_{n,k}(O)W}T}^2}
 =\frac{b}{h_{n,k}}\norm{\mathrm F}{T}^2\mper
\end{equation}
The subspace $W$ need not be invariant.  For a fixed $S\in E_{n,k}$,
the rank-one case is
\begin{equation}\label{eq:hsp-harmonic-rank-one-average}
 \int_K (U_{n,k}(O)S)(U_{n,k}(O)S)^*\,\mathrm d\mu(O)
 =\frac{\norm{\mathrm F}{S}^2}{h_{n,k}}I_{E_{n,k}}\mper
\end{equation}
Here an outer product of tensors denotes an operator after identifying
the tensor space with its coordinate space.
\end{fact}
The average in \cref{eq:hsp-harmonic-operator-average} commutes with
all rotations; irreducibility makes it scalar, and the trace determines
the scalar.  A detailed proof of all three identities is given in
\cref{app:hsp-orbit-proof}.

\begin{remark}[The harmonic restriction and the spherical orbit]
\label{rem:hsp-orbit-mechanism}
Let $\pi_{n,k}$ be the Hermitian orthogonal projection onto $E_{n,k}$,
and put $\bfxi_k(\bfx)=\pi_{n,k}\bfx^{\otimes k}$.
Its rotation orbit is
$U_{n,k}(O)\bfxi_k(\bfe_1)=\bfxi_k(O\bfe_1)$.
The pushforward of Haar probability measure under $O\mapsto O\bfe_1$
is $\sigma$.  Consequently,
\cref{eq:hsp-harmonic-rank-one-average} explains why
$\int\bfxi_k\bfxi_k^*\,\mathrm d\sigma$ is a scalar operator on
$E_{n,k}$.  The spherical Wick calculation determines its scalar as
$k!/\mathsf R_{n,k}$; the full identity is recorded in
\cref{eq:hsp-moment-operator-trace-free} in the appendix.

The harmonic restriction is essential: the full symmetric tensor
space contains different trace components and is not generally
irreducible.  Its spherical moment operator need not be scalar.
For example, for a real symmetric matrix $A$,
\[
 \int(\bfx^\top A\bfx)^2\,\mathrm d\sigma(\bfx)
 =\frac{2\norm{\mathrm F}{A}^2+(\operatorname{tr}A)^2}{n(n+2)}\mper
\]
The trace-free case recovers \cref{eq:hsp-tensor-sphere} at $k=2$.
The orbit interpretation and the quadratic calculation are justified
in \cref{app:hsp-orbit-proof}.
\end{remark}

A standard circular complex Gaussian vector $\bfZ\in\CC^m$ has independent
coordinates $(G_j+\mathrm iG'_j)/\sqrt2$, where all $G_j,G'_j$ are
independent standard real Gaussians.  If $p$ is a homogeneous
holomorphic polynomial of degree $k$ with symmetric coefficient tensor
$S$, complex Gaussian contraction gives
\begin{equation}\label{eq:hsp-complex-isometry}
 \Ex{\bfZ}{\abs{p(\bfZ)}^2}=k!\norm{\mathrm F}{S}^2\mper
\end{equation}
If $\bfZ_1,\ldots,\bfZ_k$ are independent such vectors and $S$ is evaluated
as a $k$-linear form, then
\begin{equation}\label{eq:hsp-decoupled-isometry}
 \Ex{}{\abs{S(\bfZ_1,\ldots,\bfZ_k)}^2}=\norm{\mathrm F}{S}^2\mper
\end{equation}
Both formulas follow by expanding the polynomial and using
$\Ex{}{Z_i\overline{Z_j}}=\delta_{ij}$ and
$\Ex{}{Z_iZ_j}=0$.

\subsection{A standard holomorphic Gaussian moment estimate}
\label{subsec:hsp-holomorphic-moments}

We recall a standard consequence of polarization and Gaussian decoupling.
For general decoupling background, see \cite{dPG99}.  The polarization
identity used below is the complex-phase analogue of the identity used
by Bhattiprolu et al.\ in \cite[Lemma 5.6]{BGGLT17}, where it is
attributed to He, Li, and Zhang \cite{HLZ10}.  We include the short
proof to record the circular complex Gaussian normalization and the
explicit degree-dependent constant; no novelty is claimed for this
preliminary estimate.

\begin{lemma}[A lower bound on holomorphic Gaussian moments]
\label{lem:hsp-complex-moments}
Let $p$ be a homogeneous holomorphic polynomial of degree $d$ on
$\CC^m$, and let $\bfZ$ be a standard circular complex Gaussian.
For every integer $r\ge1$,
\begin{equation}\label{eq:hsp-complex-moments}
 \Ex{}{\abs{p(\bfZ)}^{2r}}
 \ge r!\left(\frac{d!}{d^d}\right)^r
                   \left(\Ex{}{\abs{p(\bfZ)}^2}\right)^r\mper
\end{equation}
\end{lemma}
\begin{proof}
Write $p(\bfz)=S(\bfz,\ldots,\bfz)$ with $S$ a symmetric $d$-linear form.
Let $\bfZ_1,\ldots,\bfZ_d$ be independent standard circular complex Gaussian
vectors, and let $\zeta_1,\ldots,\zeta_d$ be independent uniform
phases; write $\bfzeta=(\zeta_1,\ldots,\zeta_d)$.  Extracting the
coefficient of $\zeta_1\cdots\zeta_d$ gives
\begin{equation}\label{eq:hsp-polarization}
 \Ex{\bfzeta}{\overline{\zeta_1\cdots\zeta_d}\,
       p\left(d^{-1/2}\sum_{j=1}^d\zeta_j\bfZ_j\right)}
 =d!d^{-d/2}S(\bfZ_1,\ldots,\bfZ_d)\mper
\end{equation}
For every fixed phase vector, the argument of $p$ on the left is a
standard circular complex Gaussian.  Minkowski's inequality in
$L^{2r}$ therefore yields
\[
 \norm{L^{2r}}{p(\bfZ)}
 \ge d!d^{-d/2}\norm{L^{2r}}{S(\bfZ_1,\ldots,\bfZ_d)}\mper
\]
Conditionally on $\bfZ_2,\ldots,\bfZ_d$, the multilinear form is a centered
circular complex Gaussian with some random variance $V\ge0$.
Its conditional $2r$th absolute moment is $r!V^r$.  Jensen's inequality
for this genuine probability distribution gives
\[
 \Ex{}{\abs{S(\bfZ_1,\ldots,\bfZ_d)}^{2r}}
 =r!\Ex{}{V^r}\ge r!(\Ex{}{V})^r
 =r!\norm{\mathrm F}{S}^{2r}\mper
\]
Now use \cref{eq:hsp-complex-isometry} to replace
$\norm{\mathrm F}{S}^2$ by $\Ex{}{\abs{p(\bfZ)}^2}/d!$.
\end{proof}

\subsection{Hypercontractivity and bounds for nonnegative tests}
\label{subsec:hsp-hypercontractivity}

We use the standard Gaussian polynomial hypercontractivity estimate
\cite[Eq.~(7)]{GJW22}:
\begin{equation}\label{eq:hsp-gaussian-hc}
 \norm{L^p(\gamma_m)}{G}
 \le(p-1)^{k/2}\norm{L^2(\gamma_m)}{G}
 \qquad(p\ge2,\ \deg G\le k)\mper
\end{equation}
For homogeneous $G$ this transfers to the uniform probability measure
$\sigma_m$ on $\SS^{m-1}$:
\begin{equation}\label{eq:hsp-spherical-hc}
 \norm{L^p(\sigma_m)}{G}
 \le(p-1)^{k/2}\norm{L^2(\sigma_m)}{G}\mper
\end{equation}
Indeed, write a standard real Gaussian as $R\bfX$ with independent radius
and direction.  Applying \cref{eq:hsp-gaussian-hc} leaves the radial
factor
$\Ex{}{R^{2k}}^{1/2}/\Ex{}{R^{pk}}^{1/p}\le1$.
Only homogeneity is needed for this transfer.  In particular, the
estimate applies to the degree-$r$ data polynomial $Z_g$ below
without a harmonicity assumption in $\bfa$.

\begin{lemma}[An elementary nonnegative-test estimate]
\label{lem:hsp-test-l2}
If $g\in\RR[\bfx]_{\le q}$ is nonnegative on the sphere and
$\mu_g=\int g\,\mathrm d\sigma$, then
\begin{equation}\label{eq:hsp-test-l2}
 \norm{L^2(\sigma)}{\Pi_qg}\le3^q\mu_g\mper
\end{equation}
\end{lemma}
\begin{proof}
Replace $g$ by $(g(\bfx)+g(-\bfx))/2$.  This preserves its mean and
$\Pi_qg$, and preserves nonnegativity.  Homogenize its even-degree
components to degree $q$ with powers of $\norm{\bfx}^2$; the resulting
homogeneous polynomial agrees with $g$ on the sphere.  By
\cref{eq:hsp-spherical-hc} and interpolation,
\[
 \norm{L^4(\sigma)}{g}\le3^{q/2}\norm{L^2(\sigma)}{g},
 \qquad
 \norm{L^2(\sigma)}{g}
 \le\norm{L^1(\sigma)}{g}^{1/3}
       \norm{L^4(\sigma)}{g}^{2/3}\mper
\]
Since $g\ge0$, rearrangement gives
$\norm{L^2(\sigma)}{g}\le3^q\mu_g$.
Orthogonal projection is an $L^2$ contraction.  If $\mu_g=0$,
continuity and nonnegativity imply that $g$ vanishes on the sphere.
\end{proof}

\subsection{The standard bound for the true random optimum}
\label{subsec:hsp-true-optimum}

\begin{lemma}[Coefficient-sphere concentration and a norming net]
\label{lem:hsp-true-optimum}
For every $\bfv\in\RR^h$ and $t\in\RR$,
\begin{equation}\label{eq:hsp-linear-mgf}
 \Ex{\bfa\sim\nu}{\exp(t\sqrt h\innerprod{\bfa}{\bfv})}
 \le\exp(t^2\norm{\bfv}^2/2)\mper
\end{equation}
For every $0<\epsilon<1$,
\begin{equation}\label{eq:hsp-supremum}
 \Pr{\bfa\sim\nu}{\norm{\infty}{P_{\bfa}}>
 2\sqrt{2\bigl(n\log(1+8d)+\log(2/\epsilon)\bigr)}}\le\epsilon\mper
\end{equation}
In particular, for fixed $d$, there is $C_d<\infty$ such that
$\norm{\infty}{P_{\bfa}}\le C_d\sqrt n$ with probability at least $1-e^{-n}$.
\end{lemma}
\begin{proof}
The exact spherical moment formula is
\[
 \Ex{\bfa}{\innerprod{\bfa}{\bfv}^{2k}}
   =\norm{\bfv}^{2k}\frac{(2k-1)!!}{\mathsf R_{h,k}},
 \qquad \Ex{\bfa}{\innerprod{\bfa}{\bfv}^{2k+1}}=0\mper
\]
It follows either by rotation invariance and Gaussian radius-direction
factorization or directly from spherical moments.  Since
$\mathsf R_{h,k}\ge h^k$, summing the exponential series proves
\cref{eq:hsp-linear-mgf}.  In particular, \cref{eq:hsp-basic-normalizations}
gives $\Pr{}{\abs{P_{\bfa}(\bfx)}>t}\le2e^{-t^2/2}$ at each fixed $\bfx$.

The trigonometric Bernstein inequality states
$\sup|T'|\le d\sup|T|$ for a trigonometric polynomial of degree at
most $d$; see \cite{QZ19}.  Apply it to restrictions of $P_{\bfa}$
to great circles.  A Euclidean $1/(4d)$-net of the sphere, of
cardinality at most $(1+8d)^n$, is therefore a factor-two norming set.
Indeed, its geodesic covering radius is at most $1/(2d)$, so the
variation from a maximum point to the net is at most half the
supremum.  A union bound on the net gives
\[
 \Pr{}{\norm{\infty}{P_{\bfa}}>2t}
 \le2(1+8d)^ne^{-t^2/2}\mper
\]
This proves \cref{eq:hsp-supremum}.  Take $\epsilon=e^{-n}$ for the
last assertion.
\end{proof}

%% file: sa_harmonic_lb.tex
\section{Sherali--Adams-type lower bounds}
\label{sec:hsp-contribution}

We prove \cref{thm:hsp-gap} by constructing a signed moment functional
for the powered objective.  The proof outline closely follows the
corresponding polynomial argument: retain the top spherical-harmonic
component of the powered objective, use it as a calibration direction,
control a prescribed finite family of inequalities by its covariance and
hypercontractivity, and finally lower-bound the objective by showing that
the retained top harmonic has substantial $L^2$ energy.  The main new
point needed for the last step is an explicit trace-free tensor argument:
an isotropic complex restriction kills every trace correction, while
Schur averaging guarantees that some rotated isotropic restriction retains
the correct fraction of the harmonic tensor energy.  We give that
derivation in \cref{app:hsp-isotropic-trace-energy} rather than treating
it as a standard black box.

Throughout this section, $r=q/d\in2\NN$, $h=h_{n,d}$, and
$H=h_{n,q}$, and the input model and moment region are those of
\cref{subsec:hsp-relaxation}.

For every coefficient vector $\bfa\in\SS^{h-1}$ and every real
parameter $\alpha\ge0$, define
\begin{equation}\label{eq:hsp-witness}
 \begin{aligned}
 F_{\bfa}&\defeq\Pi_q(P_{\bfa}^r),\\
 \calL_{\bfa,\alpha}(f)
 &\defeq\int f(\bfx)\bigl(1+\alpha F_{\bfa}(\bfx)\bigr)
                       \,\mathrm d\sigma(\bfx)
 \qquad(f\in\RR[\bfx]_{\le q})\mper
 \end{aligned}
\end{equation}
Here $\Pi_q$ is the spherical $L^2$ projection onto $\calH_{n,q}$.
The weight in \cref{eq:hsp-witness} may have either sign.  Nevertheless,
$F_{\bfa}$ has mean zero, and integration is supported on the sphere, so
\begin{equation}\label{eq:hsp-witness-affine}
 \calL_{\bfa,\alpha}(1)=1,\qquad
 \calL_{\bfa,\alpha}\bigl((\norm{\bfx}^2-1)f\bigr)=0
 \quad(\deg f\le q-2)
\end{equation}
for every $\alpha\ge0$.  Moreover, orthogonality gives
\begin{equation}\label{eq:hsp-exact-objective}
 \begin{aligned}
 \calL_{\bfa,\alpha}(P_{\bfa}^r)
 &=\int P_{\bfa}^r\,\mathrm d\sigma
       +\alpha\norm{L^2(\sigma)}{F_{\bfa}}^2\\
 &\ge\alpha\norm{L^2(\sigma)}{F_{\bfa}}^2\mper
 \end{aligned}
\end{equation}
The final inequality uses even $r$.  Thus a larger $\alpha$ improves
the objective; its admissible size is determined by the simultaneous
inequality constraints.  We analyze these constraints directly, without
requiring a planted distribution or positivity of the full moment matrix.

\subsection{The main lemma}
\label{subsec:hsp-main-lemma}

Our main lemma determines a feasible value of the parameter in
\cref{eq:hsp-witness} for an arbitrary prescribed finite family.

\begin{lemma}[Finite-family powered-objective lower bound]
\label{lem:hsp-main}
\label{thm:hsp-main} 
Assume $n\ge3$, $d\ge2$ is even, $2d\divides q$, and $2d\le q\le n$.
Let $\calG=\{g_1,\ldots,g_M\}$, $M\ge1$, be a family of nonnegative
polynomials on the sphere of degree at most $q$, fixed independently
of $\bfa$.  For $0<\delta<1$, set
\begin{equation}\label{eq:hsp-alpha}
 s\defeq\max\{2,\log(M/\delta)\},\qquad
 \alpha_\star\defeq
 \frac{\sqrt H}{2e\,3^q\sqrt{r!}(s-1)^{r/2}}\mper
\end{equation}
With probability at least $1-\delta$ over $\bfa\sim\nu$, the functional
$\calL_{\bfa,\alpha_\star}$ defined in \cref{eq:hsp-witness} belongs to
$\calF_{q,\calG}$.  On the same event,
\begin{equation}\label{eq:hsp-main-bound}
 \operatorname{PowLP}_{q,\calG}(P_{\bfa})
 \ge c_d\left(\frac nq\right)^{d/2}\sqrt{\frac r s},
 \qquad
 c_d\defeq\frac{d!}{54^d d^d e\sqrt2}\mper
\end{equation}
The numerical constant is not optimized.
\end{lemma}

We prove the lemma after establishing the covariance and objective
estimates in \cref{lem:hsp-covariance,lem:hsp-top-norm}.
The deduction of the main theorem appears immediately after the
lemma's proof in \cref{subsec:hsp-main-proof}.

\subsection{Data-degree structure and the exact covariance}
\label{subsec:hsp-covariance}

The projection defining $F_{\bfa}$ is in the \emph{solution variable}
$\bfx$.  Its coefficients are polynomials in the data vector $\bfa$;
we need only their degree, not a harmonic decomposition in the data
variables.  In particular, the concentration argument below uses
hypercontractivity for arbitrary homogeneous polynomials on the
coefficient sphere.

\begin{lemma}[Data homogeneity and the top-block covariance]
\label{lem:hsp-covariance}
For every fixed solution polynomial $g$, the function
\[
 Z_g(\bfa)\defeq\int g(\bfx)F_{\bfa}(\bfx)\,\mathrm d\sigma(\bfx)
\]
is a homogeneous polynomial of degree $r$ in $\bfa$ (possibly zero).
Moreover,
\begin{equation}\label{eq:hsp-covariance}
 \Ex{\bfa\sim\nu}{F_{\bfa}(\bfx)F_{\bfa}(\bfy)}
 =v_{n,d,r}Q_q^{(n)}(\innerprod{\bfx}{\bfy}),
 \quad
 v_{n,d,r}\defeq
 r!\frac{h^r}{\mathsf R_{h,r}}
       \frac{\ell_{n,d}^r}{\ell_{n,q}}
 \le r!\mper
\end{equation}
Consequently,
\begin{equation}\label{eq:hsp-test-variance}
 \Ex{\bfa\sim\nu}{Z_g(\bfa)^2}
 =\frac{v_{n,d,r}}H\norm{L^2(\sigma)}{\Pi_qg}^2\mper
\end{equation}
\end{lemma}
\begin{proof}
By self-adjointness of the solution-space projection $\Pi_q$,
\[
 Z_g(\bfa)
 =\int (\Pi_qg)(\bfx)P_{\bfa}(\bfx)^r\,\mathrm d\sigma(\bfx)
 =\sum_{i_1,\ldots,i_r=1}^h
       a_{i_1}\cdots a_{i_r}
       \int (\Pi_qg)(\bfx)
                  \prod_{t=1}^rY_{i_t}(\bfx)\,\mathrm d\sigma(\bfx)\mper
\]
The integrals are fixed scalar coefficients, so this is a degree-$r$
homogeneous polynomial in $\bfa$, with no harmonicity requirement.

For $\bfx,\bfy\in\SS^{n-1}$, spherical moment contraction in the data
variables gives
\[
 \Ex{\bfa}{P_{\bfa}(\bfx)^rP_{\bfa}(\bfy)^r}
 =\frac{h^r}{\mathsf R_{h,r}}
   \left(r!Q_d^{(n)}(\innerprod{\bfx}{\bfy})^r
        +\sum_{j\ge1}c_jQ_d^{(n)}(\innerprod{\bfx}{\bfy})^{r-2j}\right).
\]
The sum runs over $1\le j\le\lfloor r/2\rfloor$; its coefficients
are pairing counts.  The $r!$ pairings joining every $\bfx$-feature to a
$\bfy$-feature give the first term.  All remaining terms have solution
degree below $q$ and vanish under $\Pi_q$ in either variable.
Comparing leading coefficients shows that the degree-$q$ component
of $Q_d^{(n)}(t)^r$ is
$(\ell_{n,d}^r/\ell_{n,q})Q_q^{(n)}(t)$.
This proves \cref{eq:hsp-covariance}.  Its upper bound follows from
$\mathsf R_{h,r}\ge h^r$ and \cref{eq:hsp-beta-bound}.
Finally $HQ_q^{(n)}$ is the reproducing kernel for $\calH_{n,q}$;
integrating twice proves \cref{eq:hsp-test-variance}.
\end{proof}

\subsection{The top harmonic component of the powered objective}
\label{subsec:hsp-complex-polarization}

We apply the standard holomorphic Gaussian moment estimate from
\cref{lem:hsp-complex-moments} to a complex isotropic restriction of
$P$.  The following deterministic estimate is the application-specific
step that supplies the powered objective.

\begin{lemma}[Deterministic size of the top harmonic power]
\label{lem:hsp-top-norm}
Let $P\in\calH_{n,d}$ be real with $\norm{L^2(\sigma)}{P}=1$, and
put $F=\Pi_{dr}(P^r)$.  Write
$m=\lfloor n/2\rfloor$ and $b_{m,d}=\binom{m+d-1}{d}$.
For every integer $r\ge1$,
\begin{equation}\label{eq:hsp-top-norm-exact}
 \norm{L^2(\sigma)}{F}^2
 \ge \frac{r!}{\mathsf R_{n,dr}}
     \left(\frac{d!}{d^d}\right)^r
     \left(\frac{\mathsf R_{n,d}b_{m,d}}{h_{n,d}}\right)^r\mper
\end{equation}
In particular, if $d\le dr=q\le n$, then
\begin{equation}\label{eq:hsp-top-norm-simple}
 \norm{L^2(\sigma)}{F}^2
 \ge r!\left(\frac{d!}{18^d d^d}\right)^r\mper
\end{equation}
\end{lemma}
\begin{proof}
The exact estimate is the trace-free restriction calculation proved in
\cref{prop:hsp-trace-energy}; with $q=dr$, its conclusion is precisely
\cref{eq:hsp-top-norm-exact}.  We briefly recall the mechanism.
The tensor $T_P$ is first restricted to a rotated paired isotropic
subspace $V\CC^m\subseteq\CC^n$.  Schur averaging on the irreducible
harmonic tensor space guarantees a rotation for which the restriction
retains a $b_{m,d}/h_{n,d}$ fraction of the Frobenius energy.  Since
$V^\top V=0$, every vector in the image of $V$ lies on the complex null
cone, so the Fischer trace terms vanish there and
$\Pi_q(P^r)(V\bfz)=P(V\bfz)^r$.  The holomorphic Gaussian moment bound
then converts the retained degree-$d$ energy into degree-$q$ energy,
and the restriction map is contractive.  This gives
\cref{eq:hsp-top-norm-exact}; see
\cref{app:hsp-isotropic-trace-energy} for the full tensor calculation.

It remains only to simplify the constants.  Since $n\ge3$,
$m=\lfloor n/2\rfloor\ge n/3$.  Since $d\le n$,
\[
 b_{m,d}\ge\frac{(n/3)^d}{d!},\qquad
 h_{n,d}\le\binom{n+d-1}{d}\le\frac{(2n)^d}{d!}\mper
\]
Consequently $b_{m,d}/h_{n,d}\ge6^{-d}$.  Also
$\mathsf R_{n,d}\ge n^d$ and, for $q\le n$,
$\mathsf R_{n,q}\le(3n)^q$.  Substituting these inequalities into
\cref{eq:hsp-top-norm-exact} gives
\cref{eq:hsp-top-norm-simple}.  The rotated isotropic frame is only an
analytical device used to certify the norm of $F$; the witness
\cref{eq:hsp-witness} itself does not depend on this choice.
\end{proof}

\subsection{Proof of the main lemma and main theorem}
\label{subsec:hsp-main-proof}

\begin{proofof}{\cref{lem:hsp-main}}
Normalization and the sphere equalities hold for every $\alpha$ by
\cref{eq:hsp-witness-affine}.  It remains to choose a deterministic
$\alpha$ that preserves the fixed inequalities while keeping the
powered objective large.

Fix a test $g$ with $\mu_g=\int g\,\mathrm d\sigma>0$.
By \cref{lem:hsp-covariance,lem:hsp-test-l2},
\[
 \norm{L^2(\nu)}{Z_g}
 \le 3^q\sqrt{\frac{r!}{H}}\,\mu_g\mper
\]
Because $Z_g$ is homogeneous of data degree $r$, spherical
hypercontractivity in ambient dimension $h$ gives
\[
 \norm{L^s(\nu)}{Z_g}
 \le(s-1)^{r/2}3^q\sqrt{\frac{r!}{H}}\,\mu_g\mper
\]
Thus the sufficient condition
\[
 \alpha\,(s-1)^{r/2}3^q\sqrt{\frac{r!}{H}}
 \le\frac1{2e}
\]
ensures $\norm{L^s(\nu)}{\alpha Z_g}\le\mu_g/(2e)$ for every
nonnegative test.  Taking equality determines
$\alpha=\alpha_\star$ in \cref{eq:hsp-alpha}; we use this value for
the remainder of the proof.  Markov's inequality yields
\begin{equation}\label{eq:hsp-test-tail}
 \Pr{\bfa}{\abs{\alpha Z_g(\bfa)}>\mu_g/2}
 \le e^{-s}\mper
\end{equation}
A zero-mean nonnegative test vanishes on the sphere, so is satisfied
identically.  A union bound over the other tests has failure
probability at most $Me^{-s}\le\delta$.  On its complement,
\[
 \calL_{\bfa,\alpha}(g_j)=\mu_j+\alpha Z_{g_j}(\bfa)\ge\mu_j/2\ge0
 \quad(j\in[M])\mper
\]
This proves feasibility.

For every $\bfa$, independently of the test event,
\cref{eq:hsp-exact-objective,lem:hsp-top-norm} give
\begin{equation}\label{eq:hsp-root-before-simplifying}
 \calL_{\bfa,\alpha}(P_{\bfa}^r)^{1/r}
 \ge
 \frac{d!}{18^d d^d}
 \frac{H^{1/(2r)}(r!)^{1/(2r)}}
      {(2e)^{1/r}3^d\sqrt{s-1}}\mper
\end{equation}
Use \cref{eq:hsp-large-dimension},
$r!\ge(r/e)^r$, $(2e)^{1/r}\le\sqrt{2e}$ (since $r\ge2$),
and $s-1\le s$.  The resulting constant is exactly the $c_d$ in
\cref{eq:hsp-main-bound}.
\end{proofof}

\begin{proofof}{\cref{thm:hsp-gap}}
Apply \cref{lem:hsp-main} with $\delta=n^{-B}$.  Since $r=q/d$ and
$M\le n^{Cq}$, the quantity $s$ in \cref{eq:hsp-alpha} satisfies
$s\le C'_{d,C,B}q\log n$.  Therefore, with probability at least
$1-n^{-B}$,
\[
 \operatorname{PowLP}_{q,\calG}(P_{\bfa})
 \ge c'_{d,C,B}\frac{(n/q)^{d/2}}{\sqrt{\log n}}\mper
\]
Intersect this event with the event in \cref{lem:hsp-true-optimum}.
The latter has probability at least $1-e^{-n}$ and gives
$\norm{\infty}{P_{\bfa}}\le C_d\sqrt n$.  Dividing and using
\[
 n^{-1/2}(n/q)^{d/2}
 =q^{-1/2}(n/q)^{d/2-1/2}
\]
proves \cref{eq:hsp-gap} with the stated failure probability.
\end{proofof}

\begin{remark}[Range and uniformity]\label{rem:hsp-uniformity}
The feasibility argument holds for every finite fixed family of valid
tests and every even $q=dr$, without $q\le n$.  That condition is only
used to simplify the deterministic objective estimate; outside that
range, \cref{eq:hsp-top-norm-exact} and the exact dimension $H$ still
apply.  A random test family independent of $\bfa$ is also allowed by
conditioning on it.  The high-probability event is for each prescribed
family; it is not a simultaneous event over all possible families.
\end{remark}

\subsection{SA-type products and polynomial feasible-point lifts}
\label{subsec:hsp-lifts}

Let $u_1,\ldots,u_N$ be fixed nonconstant polynomials nonnegative on
the sphere.  An SA-type product family consists of the tests
\begin{equation}\label{eq:hsp-product-family}
 g_{\bfbeta}(\bfx)=\prod_{j=1}^N u_j(\bfx)^{\beta_j},
 \qquad \bfbeta\in\NN^N,\qquad
 \sum_j\beta_j\deg(u_j)\le q\mper
\end{equation}
This is the multiplication-and-linearization mechanism of the
Sherali--Adams/reformulation-linearization hierarchy introduced by
Sherali \etal~\cite{SA90}.  The generators specify the particular continuous
sphere relaxation; there is no claim that the sphere has a unique
canonical finite SA presentation.  Constants may be discarded.
There are at most $\binom{N+q}{q}$ tests.  If $N\le n^A$ and $q\le n$,
this is $n^{O_A(q)}$, so \cref{thm:hsp-gap} applies.

For a bounded explicit example, take generators $1+x_i$ and $1-x_i$,
and include all their products of degree at most $q$.
There are at most $\binom{2n+q}{q}$ tests.  They imply
$\abs{\calL(\bfx^{\bfbeta})}\le1$ for $\abs{\bfbeta}\le q$:
write $x_i$ as the difference of $(1+x_i)/2$ and $(1-x_i)/2$,
expand the product with repetitions, and use nonnegativity of all
expanded terms and the fact that their sum has functional value one.
The moment region is consequently bounded and closed.  Sphere
equalities are imposed in addition to these valid box inequalities.

\begin{proposition}[Data-independent polynomial lifts]
\label{prop:hsp-polynomial-lifts}
Let $\bfphi:\SS^{n-1}\to\RR^R$ have coordinates represented by
polynomials of degree at most $q$.  Suppose a fixed polyhedron
\[
 Q=\{\bfz:B\bfz\le \bfb,\ E\bfz=\bfe\}
\]
contains $\bfphi(\bfx)$ for every $\bfx$ on the sphere, and has $M$ inequalities.
Suppose the powered objective is represented as
\begin{equation}\label{eq:hsp-lifted-objective}
 P_{\bfa}(\bfx)^r=c_0(\bfa)+\sum_{i=1}^R c_i(\bfa)\phi_i(\bfx)
 \qquad(\bfx\in\SS^{n-1})\mper
\end{equation}
Then the $r$th root of the lifted LP optimum satisfies
\cref{eq:hsp-main-bound} with probability at least $1-\delta$.
\end{proposition}
\begin{proof}
Pull back the inequalities as
$g_j(\bfx)=b_j-(B\bfphi(\bfx))_j\ge0$, and apply \cref{lem:hsp-main}.
Set $z_i=\calL_{\bfa,\alpha_\star}(\phi_i)$.  The inequalities hold by feasibility.
The equalities hold because their pullbacks vanish on the sphere and
the constructed functional annihilates every such polynomial.
Apply the same observation to \cref{eq:hsp-lifted-objective}:
the lifted objective equals $\calL_{\bfa,\alpha_\star}(P_{\bfa}^r)$.
\end{proof}

The coefficients $c_i(\bfa)$ need not be linear in $\bfa$.  In a monomial
moment lift for the powered objective they are naturally of degree
$r$ in $\bfa$.  The data-independence assumption concerns the feasible
polyhedron and the feasible-point lift, not the objective map.
This result does not cover arbitrary extended formulations whose
feasible-point lifts may be nonpolynomial, nor tests selected after
observing $\bfa$.

\subsection{Future directions}
\label{subsec:hsp-future-directions}
\label{subsec:hsp-scope}

\paragraph{Sharp dependence on the hierarchy degree and formulation size.}
The bound in \cref{thm:hsp-gap} leaves open the optimal dependence on
$q$ and on the number of prescribed inequalities.  The finite-family
estimate \cref{eq:hsp-main-bound} depends on the test family through
$s=\max\{2,\log(M/\delta)\}$; for $M\le n^{Cq}$, this leads to the
factor $1/\sqrt{q\log n}$ in \cref{eq:hsp-gap}.  It would be useful
to determine whether this loss is intrinsic to the relaxation class,
or whether a different witness or a sharper analysis can improve it.
For particular SA-type product families, the algebraic relations among
the tests may permit a stronger simultaneous feasibility estimate than
a union bound.  Matching upper bounds for specific generator families
would help distinguish limitations of the hierarchy from losses in
the present construction.

\paragraph{Positive-semidefinite moment constructions.}
A separate direction is to extend the harmonic approach to
sum-of-squares lower bounds.  Feasibility against a prescribed finite
family of tests does not imply positivity on all degree-$q/2$ squares.
Any PSD construction must also have an objective scale compatible with
the deterministic SDP guarantee in \cref{eq:hsp-worst-case-sdp}; in
particular, it cannot simply preserve the present LP value in the
separation regime of \cref{eq:hsp-lp-sdp-separation}.

%% file: acknowledgement.tex
\section*{Acknowledgements}
\phantomsection
\addcontentsline{toc}{section}{Acknowledgements}\label{sec:ack}
The author thanks Saugata Basu for suggesting the use of spherical harmonics in the study of this
problem. The author also thanks him for explaining the fundamentals of spherical
harmonics and providing valuable insights. The author also thanks
Madhur Tulsiani and Akash Kumar for helpful technical discussions.

%% file: ai_disclosure.tex
\section*{Disclosure of AI assistance}
\phantomsection
\addcontentsline{toc}{section}{Disclosure of AI assistance}\label{sec:hsp-ai-disclosure}
The author used ChatGPT (OpenAI) during the development and preparation
of this work. The research question, the spherical-harmonic model, and
the evaluation-based planting were proposed by the author, who directed
the investigation and specified the intended relaxation framework.
The system was used to explore candidate constructions, test author suggested
mathematical estimates and proof arguments, identify potentially relevant
literature, and draft and revise portions of the exposition and the
\LaTeX{} source. This assistance was substantive and was not limited to
language editing or formatting.

%% file: appendix.tex
\section{Proofs of the spherical tensor identities}
\label{app:hsp-spherical-tensors}

We prove the spherical--tensor inner-product identity and the
orbit-averaging assertions stated in \cref{subsec:hsp-tensors}.
The auxiliary spherical Wick formula, Fischer normalization, and
harmonic moment-operator identity are stated here alongside their
proofs.
All sphere measures are probability measures, and
$\mathsf R_{n,m}=\prod_{j=0}^{m-1}(n+2j)$, with
$\mathsf R_{n,0}=1$.  The derivation proceeds from rotation
invariance to the spherical Wick formula, then to tensor
contractions and the moment-operator identity.  In particular,
the proof of \cref{lem:hsp-tensor-sphere} does not assume Gaussian
contraction rules or irreducibility.  The standard references for
the identities are Dai \etal~\cite[Theorem 1.8(3)]{DX13}
and Vignat \etal~\cite[Section 3]{VB08}.

\subsection{The spherical Wick formula}
\label{app:hsp-spherical-wick-proof}

\begin{lemma}[Spherical Wick formula]
\label{lem:hsp-spherical-wick}
Let $n\ge2$, let $m\ge0$ be an integer, and let $\sigma$ be uniform
probability measure on $\SS^{n-1}$.  Write $\calP_2(2m)$ for the
partitions of $[2m]$ into unordered pairs.  Then, for any
$i_1,\ldots,i_{2m}\in[n]$,
\begin{equation}\label{eq:hsp-spherical-pairings}
 \int x_{i_1}\cdots x_{i_{2m}}\,\mathrm d\sigma(\bfx)
 =\frac{1}{\mathsf R_{n,m}}
   \sum_{\pi\in\calP_2(2m)}
   \prod_{\{s,t\}\in\pi}\delta_{i_s,i_t}\mper
\end{equation}
Equivalently, for real vectors $\bfv_1,\ldots,\bfv_{2m}\in\RR^n$,
\[
 \int\prod_{s=1}^{2m}\innerprod{\bfv_s}{\bfx}\,\mathrm d\sigma(\bfx)
 =\frac{1}{\mathsf R_{n,m}}
   \sum_{\pi\in\calP_2(2m)}
   \prod_{\{s,t\}\in\pi}\innerprod{\bfv_s}{\bfv_t}\mper
\]
All corresponding odd moments vanish.  At $m=0$, the partition
set contains the empty partition and each empty product equals $1$.
\end{lemma}
This is the spherical Wick formula; see
Vignat \etal~\cite[Section 3]{VB08}.  We give a proof from rotation invariance and
polarization.

\begin{proof}
\emph{Step 1: rotation invariance determines the diagonal moments.}
For $m\ge0$ and $\bfv\in\RR^n$, set
\[
 A_m(\bfv)\defeq\int\innerprod{\bfv}{\bfx}^{2m}\,\mathrm d\sigma(\bfx)\mper
\]
For $m=0$ this is the constant $1$.  For $m\ge1$, the orthogonal
group acts transitively on every sphere centered at the origin, and
$\sigma$ is rotation invariant.  Hence homogeneity implies
\[
 A_m(\bfv)=a_m\norm{\bfv}^{2m}
\]
for a scalar $a_m$.  Differentiation under the integral is legitimate
because the integrand is a polynomial.  Using $\norm{\bfx}^2=1$, we get
\[
 \Delta_{\bfv} A_m(\bfv)
 =2m(2m-1)\int\norm{\bfx}^2\innerprod{\bfv}{\bfx}^{2m-2}\,\mathrm d\sigma(\bfx)
 =2m(2m-1)A_{m-1}(\bfv)\mper
\]
Direct differentiation also gives
\[
 \Delta_{\bfv}\norm{\bfv}^{2m}
 =2m(n+2m-2)\norm{\bfv}^{2m-2}\mper
\]
Comparison yields
\[
 a_m=\frac{2m-1}{n+2m-2}a_{m-1},
 \qquad a_0=1\mper
\]
Consequently,
\begin{equation}\label{eq:hsp-sphere-diagonal-moment}
 A_m(\bfv)=\frac{(2m-1)!!}{\mathsf R_{n,m}}\norm{\bfv}^{2m},
 \qquad (-1)!!\defeq1\mper
\end{equation}
This determines the normalization solely from rotation invariance
and the equation of the unit sphere.

\emph{Step 2: polarization determines all mixed moments.}
Let $\calP_2(2m)$ be the set of partitions of $[2m]$ into unordered
pairs.  There are $(2m-1)!!$ such partitions: the partner of the
first element has $2m-1$ choices, after which the count recurses.
Consider the two symmetric $2m$-linear forms
\[
 \begin{aligned}
 B_m(\bfv_1,\ldots,\bfv_{2m})
   &\defeq\int\prod_{s=1}^{2m}\innerprod{\bfv_s}{\bfx}\,\mathrm d\sigma(\bfx),\\
 C_m(\bfv_1,\ldots,\bfv_{2m})
   &\defeq\frac{1}{\mathsf R_{n,m}}
       \sum_{\pi\in\calP_2(2m)}
       \prod_{\{s,t\}\in\pi}\innerprod{\bfv_s}{\bfv_t}\mper
 \end{aligned}
\]
By \cref{eq:hsp-sphere-diagonal-moment}, the forms $B_m$ and $C_m$
agree when $\bfv_1=\cdots=\bfv_{2m}=\bfv$.  A symmetric $2m$-linear form is
uniquely determined by these diagonal values: the coefficient of
$t_1\cdots t_{2m}$ in
$B(\sum_s t_s\bfv_s,\ldots,\sum_s t_s\bfv_s)$ is
$(2m)!B(\bfv_1,\ldots,\bfv_{2m})$.  Therefore $B_m=C_m$.
Taking the $\bfv_s$ to be coordinate vectors gives
\cref{eq:hsp-spherical-pairings}.  All odd moments vanish by the
invariance under $\bfx\mapsto-\bfx$.  The case $m=0$ follows from the
empty-product convention and the normalization of $\sigma$.
\end{proof}

\subsection{The spherical--tensor inner-product identity}
\label{app:hsp-tensor-sphere-proof}

\begin{proofof}{\cref{lem:hsp-tensor-sphere}}
The case $k=0$ is immediate.  Assume $k\ge1$ and use the
spherical Wick formula in \cref{lem:hsp-spherical-wick}.

\emph{Step 1: harmonicity is exactly trace-freeness.}
For a symmetric tensor $T_B$ of order $k\ge2$, differentiating its
polynomial gives
\[
 \Delta B(\bfx)
 =k(k-1)\sum_{i_3,\ldots,i_k=1}^n
       (\operatorname{Tr}_{2}T_B)_{i_3,\ldots,i_k}
       x_{i_3}\cdots x_{i_k}\mper
\]
The tensor-to-polynomial map is injective on symmetric tensors,
as is also clear from the derivative formula for its inverse.
Thus $\Delta B=0$ if and only if $\operatorname{Tr}_{2}T_B=0$.
This holds for complex coefficients as well; conjugating a
trace-free tensor leaves it trace-free.  For $k=1$ there is no trace
condition.

\emph{Step 2: only cross-pairings survive.}
Expand $A(\bfx)\overline{B(\bfx)}$ using their tensors and apply
\cref{eq:hsp-spherical-pairings} with $m=k$.  In each pairing,
regard the first $k$ slots as belonging to $T_A$ and the last $k$
slots as belonging to $\overline{T_B}$.  If $u$ pairs are internal
to the first block, $v$ pairs are internal to the second block, and
$w$ pairs cross between blocks, then
\[
 2u+w=k=2v+w\mper
\]
Thus $u=v$.  Any pairing that is not entirely cross-block has an
internal pair in the second block.  Summing the index of that pair
contracts two slots of $\overline{T_B}$ and gives zero by
trace-freeness.  This also explains why no trace-free assumption on
$T_A$ is necessary.

The surviving pairings are the bijections between the two blocks;
there are exactly $k!$ of them.  For each such bijection, symmetry
of $T_B$ makes the resulting contraction equal to
\[
 \sum_{i_1,\ldots,i_k=1}^n
 (T_A)_{i_1,\ldots,i_k}
 \overline{(T_B)_{i_1,\ldots,i_k}}
 =\innerprod{T_A}{T_B}_{\mathrm F}\mper
\]
All pairings have the common prefactor $1/\mathsf R_{n,k}$.
This proves \cref{eq:hsp-tensor-sphere-polarized}; taking $A=B=G$
proves \cref{eq:hsp-tensor-sphere}.
\end{proofof}

\subsection{Conversion to the Fischer inner product}
\label{app:hsp-fischer-proof}

\begin{remark}[Translation to the Fischer inner product]
\label{rem:hsp-fischer-normalization}
For homogeneous degree-$k$ polynomials
$A(\bfx)=\sum_{|\bfalpha|=k}a_{\bfalpha} \bfx^{\bfalpha}$ and
$B(\bfx)=\sum_{|\bfalpha|=k}b_{\bfalpha} \bfx^{\bfalpha}$, put
$\bfalpha!=\prod_i\alpha_i!$ and
\[
 [A,B]_{\partial}\defeq
 \sum_{|\bfalpha|=k}\bfalpha!a_{\bfalpha}\overline{b_{\bfalpha}}\mper
\]
Then
\[
 k!\innerprod{T_A}{T_B}_{\mathrm F}
 =[A,B]_{\partial}=A(\partial)\overline{B}\mper
\]
If $B$ is harmonic, this becomes
\[
 [A,B]_{\partial}
 =2^k(n/2)_k\int A\overline{B}\,\mathrm d\sigma
 =\mathsf R_{n,k}\int A\overline{B}\,\mathrm d\sigma\mper
\]
This is the normalization in Dai \etal~\cite[Theorem 1.8(3)]{DX13},
with dimension and degree renamed.  The tensor norm is over
\emph{ordered} indices, not the unweighted Euclidean norm of the
monomial coefficient vector.  For unnormalized surface-area measure,
the right-hand side of \cref{eq:hsp-tensor-sphere} must be multiplied
by the surface area.
\end{remark}

\begin{proof}
Write $A(\bfx)=\sum_{|\bfalpha|=k}a_{\bfalpha} \bfx^{\bfalpha}$ and
$B(\bfx)=\sum_{|\bfalpha|=k}b_{\bfalpha} \bfx^{\bfalpha}$.  For each multi-index
$\bfalpha$ of total degree $k$, there are $k!/\bfalpha!$ ordered tuples
with multiplicity vector $\bfalpha$.  The tensor entry of $T_A$ on
each such tuple is $\bfalpha!a_{\bfalpha}/k!$, and similarly for $T_B$.
Consequently,
\[
 \begin{aligned}
 k!\innerprod{T_A}{T_B}_{\mathrm F}
 &=k!\sum_{|\bfalpha|=k}\frac{k!}{\bfalpha!}
      \frac{\bfalpha!a_{\bfalpha}}{k!}
      \frac{\bfalpha!\overline{b_{\bfalpha}}}{k!}\\
 &=\sum_{|\bfalpha|=k}\bfalpha!a_{\bfalpha}\overline{b_{\bfalpha}}
 =[A,B]_{\partial}\mper
 \end{aligned}
\]
Also, $\partial^{\bfalpha} \bfx^{\bfbeta}$ is the constant $\bfalpha!$ when
$\bfalpha=\bfbeta$ and is zero otherwise, since both total degrees are
$k$.  Thus $A(\partial)\overline B=[A,B]_{\partial}$.
When $B$ is harmonic, \cref{eq:hsp-tensor-sphere-polarized} now gives
\[
 [A,B]_{\partial}
 =\mathsf R_{n,k}\int A\overline B\,\mathrm d\sigma
 =2^k(n/2)_k\int A\overline B\,\mathrm d\sigma\mper
\]
The last equality follows by factoring $2$ from each term in
$\mathsf R_{n,k}$.  Replacing probability measure by surface-area
measure multiplies the integral by the total surface area, which
accounts for the stated change of normalization.
\end{proof}

\subsection{The moment operator on the harmonic subspace}
\label{app:hsp-moment-operator-proof}

\begin{corollary}[Moment operator on the trace-free subspace]
\label{cor:hsp-harmonic-moment-operator}
Let $E_{n,k}$ be the space of symmetric trace-free complex tensors
of order $k$, and let $\pi_{n,k}$ be its Hermitian orthogonal
projection inside $(\CC^n)^{\otimes k}$.  Define
\[
 C_{n,k}\defeq\int \bfx^{\otimes k}(\bfx^{\otimes k})^*
                       \,\mathrm d\sigma(\bfx),
 \qquad
 \bfxi_k(\bfx)\defeq\pi_{n,k}\bfx^{\otimes k},
 \qquad c_{n,k}\defeq\frac{k!}{\mathsf R_{n,k}}\mper
\]
Write $h_{n,k}=\dim E_{n,k}$.  Then
\begin{equation}\label{eq:hsp-moment-operator-trace-free}
 C_{n,k}\pi_{n,k}=\pi_{n,k}C_{n,k}=c_{n,k}\pi_{n,k},
 \qquad
 \int\bfxi_k(\bfx)\bfxi_k(\bfx)^*\,\mathrm d\sigma(\bfx)
       =c_{n,k}I_{E_{n,k}}\mper
\end{equation}
The second identity is understood as an identity of operators on
$E_{n,k}$.  Moreover,
\begin{equation}\label{eq:hsp-evaluation-orbit-norm}
 \norm{\mathrm F}{\bfxi_k(\bfx)}^2=c_{n,k}h_{n,k}
 \qquad(\bfx\in\SS^{n-1})\mper
\end{equation}
\end{corollary}

\begin{proof}
The mixed identity \cref{eq:hsp-tensor-sphere-polarized} shows that
$C_{n,k}T=c_{n,k}T$ for every $T\in E_{n,k}$, by testing against
arbitrary symmetric tensors.  The operator $C_{n,k}$ has range in
the symmetric tensors, so there is no component in their
orthogonal complement.  Its self-adjointness gives the identity
with the factors reversed.  Compressing by $\pi_{n,k}$ proves the
second assertion in \cref{eq:hsp-moment-operator-trace-free}.

The space $E_{n,k}$ is invariant under orthogonal changes of
coordinates, so its orthogonal projection commutes with them.
Hence $\norm{\mathrm F}{\bfxi_k(\bfx)}$ is constant on the unit sphere.
Taking the trace of the covariance identity gives the squared
value $c_{n,k}\dim E_{n,k}=c_{n,k}h_{n,k}$.
\end{proof}

\subsection{Orbit averaging and the trace contribution}
\label{app:hsp-orbit-proof}

\paragraph{The complex form of the quoted irreducibility theorem.}
The proof in Dai \etal~\cite[Theorem 7.2]{DX13} applies to complex invariant
subspaces as follows.  Let $W\subseteq\calH_{n,k}^{\CC}$ be invariant
and let $K_W(\bfx,\bfy)=\sum_jY_j(\bfx)\overline{Y_j(\bfy)}$ be
the kernel of its Hermitian orthogonal projection, for an orthonormal
basis of $W$.  Unitarity makes $K_W$ rotation invariant.  For fixed
$\bfy$, it is a harmonic of degree $k$ in $\bfx$ and invariant under
the stabilizer of $\bfy$.  The one-dimensional zonal-space statement
of Dai \etal~\cite[Lemma 7.1]{DX13}, valid also for complex coefficients,
implies $K_W(\bfx,\bfy)=c(\bfy)K_{n,k}(\bfx,\bfy)$, where
$K_{n,k}$ is the full harmonic reproducing kernel.  Rotation
transitivity makes $c(\bfy)$ constant.  Hence the orthogonal projection
onto $W$ is $cI$ on $\calH_{n,k}^{\CC}$.  Idempotence gives
$c\in\{0,1\}$, so $W$ is zero or the full harmonic space.
This is the complex version of the cited argument, not an inference
from real irreducibility alone.

\begin{proofof}{\cref{fact:hsp-harmonic-averaging,rem:hsp-orbit-mechanism}}
We first prove the operator identity in a general irreducible unitary
representation.  Let $U:K\to\operatorname{U}(E)$ be such a
representation of a compact group on a nonzero finite-dimensional
complex inner-product space, let $\mu_K$ be Haar probability measure,
and let $A$ be a Hermitian operator on $E$.  Define
\[
 \overline A\defeq\int_K U(g)A U(g)^*\,\mathrm d\mu_K(g)\mper
\]
The integral exists entrywise.  For every $h\in K$, left invariance
of Haar measure gives
\[
 U(h)\overline A U(h)^*
 =\int_K U(hg)A U(hg)^*\,\mathrm d\mu_K(g)
 =\overline A\mper
\]
Thus $\overline A$ commutes with the representation.  It is Hermitian,
so its eigenspaces form an orthogonal decomposition of $E$.
Every eigenspace is invariant under each $U(h)$, because
$\overline A U(h)=U(h)\overline A$.  Irreducibility implies that
there is a single eigenvalue, say $c$, and hence
$\overline A=cI_E$.  Unitarity and cyclicity of trace imply
\[
 c\dim E=\operatorname{tr}\overline A
 =\int_K\operatorname{tr}(U(g)A U(g)^*)\,\mathrm d\mu_K(g)
 =\operatorname{tr}A\mper
\]
Apply this to $E_{n,k}$ and $U_{n,k}$, which are irreducible by
\cref{thm:hsp-harmonic-irreducibility} and the tensor--polynomial
identification following that theorem.  This proves
\cref{eq:hsp-harmonic-operator-average} for either rotation group.

For the projector specialization, unitarity gives
$A_{U_{n,k}(O)W}=U_{n,k}(O)A_WU_{n,k}(O)^*$.
Since $\operatorname{tr}A_W=\dim W=b$, the averaged projector equals
$(b/h_{n,k})I_{E_{n,k}}$.  For an orthogonal projection $B$,
$\norm{\mathrm F}{BT}^2=\innerprod{BT}{T}_{\mathrm F}$.
Integrating this equality yields the second assertion in
\cref{eq:hsp-harmonic-projector-average}.  Taking $A=SS^*$, whose
trace is $\norm{\mathrm F}{S}^2$, proves
\cref{eq:hsp-harmonic-rank-one-average}.

For the spherical orbit, the invariant subspace $E_{n,k}$ and its
orthogonal complement are preserved by $O^{\otimes k}$, so
$\pi_{n,k}$ commutes with $O^{\otimes k}$.
It follows that
\[
 U_{n,k}(O)\bfxi_k(\bfe_1)
 =O^{\otimes k}\pi_{n,k}\bfe_1^{\otimes k}
 =\pi_{n,k}(O\bfe_1)^{\otimes k}
 =\bfxi_k(O\bfe_1)\mper
\]
The pushforward of Haar measure is the rotation-invariant probability
measure $\sigma$.  The rank-one averaging identity therefore gives
\[
 \int\bfxi_k(\bfx)\bfxi_k(\bfx)^*\,\mathrm d\sigma(\bfx)
 =\frac{\norm{\mathrm F}{\bfxi_k(\bfe_1)}^2}{h_{n,k}}I_{E_{n,k}}\mper
\]
By \cref{eq:hsp-evaluation-orbit-norm}, the numerator is
$h_{n,k}k!/\mathsf R_{n,k}$, in agreement with the direct
spherical-Wick computation.  This use of the tensor--sphere identity
only evaluates the scalar; it is not used to prove the quoted
irreducibility theorem.

For the quadratic example, expand
$(\bfx^\top A\bfx)^2=\sum_{i,j,k,\ell}A_{ij}A_{k\ell}x_ix_jx_kx_\ell$.
The three pairings in \cref{eq:hsp-spherical-pairings} at $m=2$
give
\[
 \begin{aligned}
 n(n+2)\int(\bfx^\top A\bfx)^2\,\mathrm d\sigma
 &=\sum_{i,j,k,\ell}A_{ij}A_{k\ell}
      (\delta_{ij}\delta_{k\ell}
       +\delta_{ik}\delta_{j\ell}
       +\delta_{i\ell}\delta_{jk})\\
 &=(\operatorname{tr}A)^2+2\norm{\mathrm F}{A}^2\mper
 \end{aligned}
\]
Here symmetry of $A$ identifies each of the last two sums with
$\norm{\mathrm F}{A}^2$.  Finally,
$\Delta(\bfx^\top A\bfx)=2\operatorname{tr}A$, so the additional trace term
vanishes precisely for a harmonic quadratic form.
\end{proofof}

\subsection{Isotropic restrictions and trace-free energy}
\label{app:hsp-isotropic-trace-energy}

We give here the application-specific tensor calculation used in
\cref{lem:hsp-top-norm}.  The representation-theoretic input is only the
orbit-averaging identity from \cref{fact:hsp-harmonic-averaging}; the
remaining statements are elementary consequences of isotropy and the
Fischer decomposition.  We include the details because the exact
trace-free restriction statement, in the normalization needed here, is
not being invoked from an external theorem.

Put $m=\lfloor n/2\rfloor$ and define the paired isotropic embedding
$V_0:\CC^m\to\CC^n$ by
\begin{equation}\label{eq:hsp-paired-isotropic-frame}
 V_0\bfz
 =\frac1{\sqrt2}(z_1,\mathrm i z_1,z_2,\mathrm i z_2,\ldots,
                  z_m,\mathrm i z_m,0,\ldots,0)\mper
\end{equation}
Thus
\begin{equation}\label{eq:hsp-frame-identities}
 V_0^*V_0=I_m,
 \qquad
 V_0^\top V_0=0\mper
\end{equation}
The same identities hold for $V=OV_0$ with $O\in O(n)$.

For a symmetric complex tensor $T$ of order $k$, write
\[
 R_VT\defeq(V^\top)^{\otimes k}T,
 \qquad
 L_VS\defeq\overline V^{\otimes k}S
 \quad(S\in\operatorname{Sym}^k(\CC^m))\mper
\]
The map $L_V$ is a Hermitian isometry because $V^*V=I_m$, and its
adjoint is $R_V$.  Hence, if $A_V$ denotes Hermitian orthogonal
projection onto
\[
 W_V\defeq
 L_V\bigl(\operatorname{Sym}^k(\CC^m)\bigr)
 =\operatorname{Sym}^k(\overline{\operatorname{im}V}),
\]
then
\begin{equation}\label{eq:hsp-restriction-projection-norm}
 \norm{\mathrm F}{R_VT}^2
 =\norm{\mathrm F}{A_VT}^2
 \le \norm{\mathrm F}{T}^2\mper
\end{equation}
Moreover, $W_V$ is contained in the trace-free tensor space.  Indeed,
contracting any two tensor slots of $\overline V^{\otimes k}S$ produces
a factor $\overline{V^\top V}=0$.  Consequently
\begin{equation}\label{eq:hsp-isotropic-subspace-dimension}
 W_V\subseteq E_{n,k},
 \qquad
 \dim_{\CC}W_V=\dim_{\CC}\operatorname{Sym}^k(\CC^m)
              =\binom{m+k-1}{k}\mper
\end{equation}

\begin{lemma}[Trace-free projection is invisible on the null cone]
\label{lem:hsp-null-cone-projection}
Let $G$ be a homogeneous polynomial of degree $k$, possibly with
complex coefficients, and let $G_k=\Pi_kG$ be its top harmonic
component.  If $\bfz\in\CC^n$ satisfies $\bfz^\top\bfz=0$, then
\begin{equation}\label{eq:hsp-null-cone-projection}
 G_k(\bfz)=G(\bfz)\mper
\end{equation}
Equivalently, if $T_G$ and $T_{G_k}$ are their symmetric coefficient
tensors, every isotropic frame $V$ satisfies
\begin{equation}\label{eq:hsp-null-restriction-tensor}
 R_VT_{G_k}=R_VT_G\mper
\end{equation}
\end{lemma}

\begin{proof}
By the Fischer decomposition,
$G(\bfx)=G_k(\bfx)+\norm{\bfx}^2R(\bfx)$ for a homogeneous polynomial
$R$ of degree $k-2$.  This algebraic identity extends to complex
arguments, with $\norm{\bfx}^2$ replaced by the quadratic polynomial
$x_1^2+\cdots+x_n^2$.  Thus the second term vanishes whenever
$\bfz^\top\bfz=0$, proving \cref{eq:hsp-null-cone-projection}.
For $\bfz\in\CC^m$, \cref{eq:hsp-frame-identities} gives
$(V\bfz)^\top(V\bfz)=0$.  The two restricted homogeneous polynomials
therefore agree identically, and equality of their symmetric
coefficient tensors gives \cref{eq:hsp-null-restriction-tensor}.
\end{proof}

\begin{lemma}[Energy retained by an isotropic restriction]
\label{lem:hsp-isotropic-restriction-energy}
Let $T\in E_{n,d}$ and $m=\lfloor n/2\rfloor$.  For
$V_O=OV_0$, $O\in O(n)$ Haar distributed, and
$b_{m,d}=\binom{m+d-1}{d}$,
\begin{equation}\label{eq:hsp-isotropic-average}
 \Ex{O}{\norm{\mathrm F}{R_{V_O}T}^2}
 =\frac{b_{m,d}}{h_{n,d}}\norm{\mathrm F}{T}^2\mper
\end{equation}
Consequently there is a rotation $O$ for which
\begin{equation}\label{eq:hsp-isotropic-retained}
 \norm{\mathrm F}{R_{V_O}T}^2
 \ge\frac{b_{m,d}}{h_{n,d}}\norm{\mathrm F}{T}^2\mper
\end{equation}
\end{lemma}

\begin{proof}
By \cref{eq:hsp-restriction-projection-norm},
$\norm{\mathrm F}{R_{V_O}T}^2=\norm{\mathrm F}{A_{V_O}T}^2$.
The subspace $W_{V_0}$ lies in $E_{n,d}$ and has dimension $b_{m,d}$
by \cref{eq:hsp-isotropic-subspace-dimension}; moreover
$W_{V_O}=U_{n,d}(O)W_{V_0}$.  The projector form of Schur averaging,
\cref{eq:hsp-harmonic-projector-average}, therefore gives
\cref{eq:hsp-isotropic-average}.  Some rotation attains at least the
mean, which proves \cref{eq:hsp-isotropic-retained}.
\end{proof}

\begin{proposition}[Trace-free energy of a powered harmonic]
\label{prop:hsp-trace-energy}
Let $P\in\calH_{n,d}$ be real with
$\norm{L^2(\sigma)}{P}=1$, let $r\ge1$, put $q=dr$, and let
$F=\Pi_q(P^r)$.  Then, with $m=\lfloor n/2\rfloor$ and
$b_{m,d}=\binom{m+d-1}{d}$,
\begin{equation}\label{eq:hsp-trace-energy}
 \norm{L^2(\sigma)}{F}^2
 \ge \frac{r!}{\mathsf R_{n,q}}
     \left(\frac{d!}{d^d}\right)^r
     \left(\frac{\mathsf R_{n,d}b_{m,d}}{h_{n,d}}\right)^r\mper
\end{equation}
\end{proposition}

\begin{proof}
Let $T=T_P\in E_{n,d}$.  The tensor--sphere identity gives
\begin{equation}\label{eq:hsp-input-tensor-energy}
 \norm{\mathrm F}{T}^2=\frac{\mathsf R_{n,d}}{d!}\mper
\end{equation}
Choose a rotated paired frame $V$ satisfying
\cref{eq:hsp-isotropic-retained}, and define
$p(\bfz)=P(V\bfz)$.  Its symmetric coefficient tensor is $R_VT$.
By the complex Gaussian isometry and
\cref{eq:hsp-isotropic-retained,eq:hsp-input-tensor-energy},
\begin{equation}\label{eq:hsp-isotropic-energy}
 \Ex{\bfZ}{\abs{p(\bfZ)}^2}
 =d!\norm{\mathrm F}{R_VT}^2
 \ge\frac{\mathsf R_{n,d}b_{m,d}}{h_{n,d}}\mper
\end{equation}

Every vector $V\bfz$ lies on the complex null cone.  Applying
\cref{lem:hsp-null-cone-projection} to $G=P^r$ therefore yields
\begin{equation}\label{eq:hsp-powered-null-restriction}
 F(V\bfz)=P(V\bfz)^r=p(\bfz)^r\mper
\end{equation}
If $T_F$ is the tensor of $F$, then the tensor of the left-hand side
is $R_VT_F$.  Hence the degree-$q$ Gaussian isometry, followed by the
contraction in \cref{eq:hsp-restriction-projection-norm} and then the
spherical--tensor identity, gives
\begin{equation}\label{eq:hsp-restricted-powered-upper}
 \Ex{\bfZ}{\abs{p(\bfZ)}^{2r}}
 =q!\norm{\mathrm F}{R_VT_F}^2
 \le q!\norm{\mathrm F}{T_F}^2
 =\mathsf R_{n,q}\norm{L^2(\sigma)}{F}^2\mper
\end{equation}
On the other hand, the holomorphic moment estimate
\cref{lem:hsp-complex-moments} and
\cref{eq:hsp-isotropic-energy} imply
\[
 \Ex{\bfZ}{\abs{p(\bfZ)}^{2r}}
 \ge r!\left(\frac{d!}{d^d}\right)^r
       \left(\frac{\mathsf R_{n,d}b_{m,d}}{h_{n,d}}\right)^r\mper
\]
Comparing with \cref{eq:hsp-restricted-powered-upper} proves
\cref{eq:hsp-trace-energy}.
\end{proof}

\section{The harmonic spectrum of the Gaussian tensor ensemble}
\label{app:hsp-kostlan-spectrum}

We verify the normalization of \cref{eq:hsp-kostlan-eigenvalues}
directly.  These are the classical Kostlan--Shub--Smale harmonic
weights; see Gichev~\cite[Proposition 3.1]{Gic17}.

Let $G$ be the Gaussian polynomial in \cref{eq:hsp-kostlan-model}.
The multinomial identity gives
$\Ex{}{G(\bfx)G(\bfy)}=(\bfx^\top\bfy)^d$.
Consequently its covariance operator on $L^2(\sigma)$ is
\[
 (\calC_{\mathrm K}f)(\bfx)
 =\int(\bfx^\top\bfy)^d f(\bfy)\,\mathrm d\sigma(\bfy)\mper
\]
Fix an admissible harmonic degree $j$ and put $k=(d-j)/2$.
For $Y\in\calH_{n,j}$, write
$Y(\bfy)=\sum_{i_1,\ldots,i_j}(T_Y)_{i_1,\ldots,i_j}y_{i_1}\cdots y_{i_j}$,
where $T_Y$ is symmetric and trace-free.  Apply the spherical Wick
formula, \cref{lem:hsp-spherical-wick}, to the $d+j$ factors of $\bfy$
in $(\bfx^\top\bfy)^dY(\bfy)$.

Any pairing internal to the $j$ slots of $T_Y$ vanishes by
trace-freeness.  Therefore each of these $j$ slots must be paired
with a distinct slot from the $d$ copies of $\bfx$.  This can be done
in $d!/(d-j)!$ ways.  The remaining $d-j=2k$ slots can be paired in
$(2k-1)!!=(2k)!/(2^k k!)$ ways.  Each surviving contraction is
$Y(\bfx)\norm{\bfx}^{2k}$, and its denominator is
$\mathsf R_{n,(d+j)/2}$.
For $\bfx\in\SS^{n-1}$ we obtain
\[
 (\calC_{\mathrm K}Y)(\bfx)
 =\frac{d!}{2^k k!\,\mathsf R_{n,(d+j)/2}}Y(\bfx),
\]
which proves \cref{eq:hsp-kostlan-eigenvalues}.
For $j=0$ the same calculation has no tensor slots; the empty-pairing
conventions also cover $k=0$.
Harmonic components of degree above $d$, or of the wrong parity,
are annihilated by the operator.

Because $G$ is Gaussian, its coefficients in an orthonormal harmonic
basis are independent centered Gaussians with the corresponding
variances $\lambda_{n,d,j}$.  Dividing consecutive values gives
\cref{eq:hsp-kostlan-ratio}.  The trace contribution of degree $j$
is $h_{n,j}\lambda_{n,d,j}$, and all contributions sum to one because
$\Ex{}{G(\bfx)^2}=1$ on the sphere.
Using the dimension and leading-coefficient identity
\cref{eq:hsp-leading-identity}, the top-level mass is
\[
 h_{n,d}\lambda_{n,d,d}
 =\frac1{\ell_{n,d}}
 =\prod_{i=0}^{d-1}\frac{n-2+i}{n-2+2i}
 =1-\frac{d(d-1)}{2n}+O_d(n^{-2})\mper
\]
This proves \cref{eq:hsp-kostlan-top-mass} and explains why
per-coefficient lower-degree bias does not imply lower-degree
concentration of total mass when $d$ is fixed.

Finally, the Hilbert--Schmidt kernel identity and spherical Wick give
\[
 \operatorname{tr}(\calC_{\mathrm K}^2)
 =\iint(\bfx^\top\bfy)^{2d}\,\mathrm d\sigma(\bfx)\mathrm d\sigma(\bfy)
 =\frac{(2d-1)!!}{\mathsf R_{n,d}}\mper
\]
Since the covariance of $P_{\bfa}$ is $\Pi_d/h$, its squared trace norm
is $1/h$.  Both covariance operators have trace one, and
$h\sim n^d/d!$ and $\mathsf R_{n,d}\sim n^d$ for fixed $d$.
These observations prove \cref{eq:hsp-effective-dimension}.

%% file: references.bib
@misc{DX13,
  author        = {Dai, Feng and Xu, Yuan},
  title         = {Spherical Harmonics},
  year          = {2013},
  eprint        = {1304.2585},
  archivePrefix = {arXiv},
  primaryClass  = {math.CA},
  note          = {Chapter 1 of Approximation Theory and Harmonic Analysis on Spheres and Balls},
  url           = {https://arxiv.org/abs/1304.2585}
}

@misc{GJW22,
  author        = {Gamarnik, David and Jagannath, Aukosh and Wein, Alexander S.},
  title         = {Hardness of Random Optimization Problems for {Boolean} Circuits, Low-Degree Polynomials, and {Langevin} Dynamics},
  year          = {2022},
  eprint        = {2004.12063},
  archivePrefix = {arXiv},
  primaryClass  = {cs.CC},
  note          = {Version 2; Gaussian polynomial hypercontractivity is stated in equation (7)},
  url           = {https://arxiv.org/abs/2004.12063v2}
}

@article{SA90,
  author        = {Sherali, Hanif D. and Adams, Warren P.},
  title         = {A Hierarchy of Relaxations between the Continuous and Convex Hull Representations for Zero-One Programming Problems},
  journal       = {SIAM Journal on Discrete Mathematics},
  year          = {1990},
  volume        = {3},
  number        = {3},
  pages         = {411--430},
  doi           = {10.1137/0403036}
}

@misc{HKP17,
  author        = {Hopkins, Samuel B. and Kothari, Pravesh K. and Potechin, Aaron and Raghavendra, Prasad and Schramm, Tselil and Steurer, David},
  title         = {The Power of Sum-of-Squares for Detecting Hidden Structures},
  year          = {2017},
  eprint        = {1710.05017},
  archivePrefix = {arXiv},
  primaryClass  = {cs.DS},
  url           = {https://arxiv.org/abs/1710.05017}
}

@misc{QZ19,
  author        = {Queff{\'e}lec, Herv{\'e} and Zarouf, Rachid},
  title         = {On {Bernstein}'s Inequality for Polynomials},
  year          = {2019},
  eprint        = {1903.10801},
  archivePrefix = {arXiv},
  primaryClass  = {math.CA},
  url           = {https://arxiv.org/abs/1903.10801}
}

@misc{BGGLT17,
  author        = {Bhattiprolu, Vijay and Ghosh, Mrinalkanti and Guruswami, Venkatesan and Lee, Euiwoong and Tulsiani, Madhur},
  title         = {Weak Decoupling, Polynomial Folds, and Approximate Optimization over the Sphere},
  year          = {2017},
  eprint        = {1611.05998},
  archivePrefix = {arXiv},
  primaryClass  = {cs.DS},
  note          = {Version 2, revised April 22, 2017; first posted November 18, 2016},
  doi           = {10.48550/arXiv.1611.05998},
  url           = {https://arxiv.org/abs/1611.05998v2}
}

@article{VB08,
  author        = {Vignat, C. and Bhatnagar, S.},
  title         = {An Extension of {Wick}'s Theorem},
  journal       = {Statistics \& Probability Letters},
  year          = {2008},
  volume        = {78},
  number        = {15},
  pages         = {2404--2407},
  doi           = {10.1016/j.spl.2008.03.003},
  eprint        = {0709.1999},
  archivePrefix = {arXiv},
  primaryClass  = {cond-mat.stat-mech},
  url           = {https://arxiv.org/abs/0709.1999}
}

@book{dPG99,
  author    = {de la Pe{\~n}a, V{\'i}ctor H. and Gin{\'e}, Evarist},
  title     = {Decoupling: From Dependence to Independence},
  series    = {Probability and Its Applications},
  publisher = {Springer},
  address   = {New York},
  year      = {1999},
  doi       = {10.1007/978-1-4612-0537-1},
  isbn      = {978-0-387-98616-6}
}

@article{HLZ10,
  author  = {He, Simai and Li, Zhening and Zhang, Shuzhong},
  title   = {Approximation Algorithms for Homogeneous Polynomial Optimization with Quadratic Constraints},
  journal = {Mathematical Programming},
  year    = {2010},
  volume  = {125},
  number  = {2},
  pages   = {353--383},
  doi     = {10.1007/s10107-010-0409-z}
}

@misc{Gic17,
  author        = {Gichev, V.},
  title         = {Decomposition of the {Kostlan--Shub--Smale} Model for Random Polynomials},
  year          = {2017},
  eprint        = {1509.01449},
  archivePrefix = {arXiv},
  primaryClass  = {math.CA},
  note          = {Version 2; Proposition 3.1 and Theorems 4.2 and 5.3},
  url           = {https://arxiv.org/abs/1509.01449v2}
}

@misc{FLL15,
  author        = {Fyodorov, Yan and Lerario, Antonio and Lundberg, Erik},
  title         = {On the Number of Connected Components of Random Algebraic Hypersurfaces},
  year          = {2015},
  eprint        = {1404.5349},
  archivePrefix = {arXiv},
  primaryClass  = {math.AG},
  note          = {Version 2; invariant ensembles and the examples of Section 8},
  url           = {https://arxiv.org/abs/1404.5349v2}
}

@misc{BGL17,
  author        = {Bhattiprolu, Vijay and Guruswami, Venkatesan and Lee, Euiwoong},
  title         = {Sum-of-Squares Certificates for Maxima of Random Tensors on the Sphere},
  year          = {2017},
  eprint        = {1605.00903},
  archivePrefix = {arXiv},
  primaryClass  = {cs.CC},
  note          = {Version 2; first posted in 2016},
  url           = {https://arxiv.org/abs/1605.00903v2}
}

@article{HL13,
  author  = {Hillar, Christopher J. and Lim, Lek-Heng},
  title   = {Most Tensor Problems Are {NP}-Hard},
  journal = {Journal of the ACM},
  year    = {2013},
  volume  = {60},
  number  = {6},
  pages   = {45:1--45:39},
  doi     = {10.1145/2512329},
  eprint  = {0911.1393},
  archivePrefix = {arXiv},
  primaryClass = {cs.CC}
}

@article{Las01,
  author  = {Lasserre, Jean B.},
  title   = {Global Optimization with Polynomials and the Problem of Moments},
  journal = {SIAM Journal on Optimization},
  year    = {2001},
  volume  = {11},
  number  = {3},
  pages   = {796--817},
  doi     = {10.1137/S1052623400366802}
}

@phdthesis{Par00,
  author = {Parrilo, Pablo A.},
  title  = {Structured Semidefinite Programs and Semialgebraic Geometry Methods in Robustness and Optimization},
  school = {California Institute of Technology},
  year   = {2000},
  doi    = {10.7907/2K6Y-CH43},
  url    = {https://thesis.caltech.edu/1647/}
}

@article{So11,
  author  = {So, Anthony Man-Cho},
  title   = {Deterministic Approximation Algorithms for Sphere Constrained Homogeneous Polynomial Optimization Problems},
  journal = {Mathematical Programming},
  year    = {2011},
  volume  = {129},
  number  = {2},
  pages   = {357--382},
  doi     = {10.1007/s10107-011-0464-0}
}

@inproceedings{HKPT24,
  author    = {Hsieh, Jun-Ting and Kothari, Pravesh K. and Pesenti, Lucas and Trevisan, Luca},
  title     = {New {SDP} Roundings and Certifiable Approximation for Cubic Optimization},
  booktitle = {Proceedings of the 2024 Annual ACM-SIAM Symposium on Discrete Algorithms (SODA)},
  year      = {2024},
  pages     = {2337--2362},
  publisher = {SIAM},
  doi       = {10.1137/1.9781611977912.83},
  eprint    = {2310.00393},
  archivePrefix = {arXiv},
  primaryClass = {cs.DS}
}

@misc{JWX26,
  author = {Jeronimo, Fernando Granha and Wu, Pei and Xu, Haochen},
  title  = {An Argmax Principle for Sum-of-Squares Relaxations on the Sphere},
  year   = {2026},
  eprint = {2608.02594},
  archivePrefix = {arXiv},
  primaryClass = {cs.CC},
  url    = {https://arxiv.org/abs/2608.02594},
  note   = {Theorem 1.4 and Section 6 recover the general degree-dependent approximation guarantee}
}

@misc{DW12,
  author = {Doherty, Andrew C. and Wehner, Stephanie},
  title  = {Convergence of {SDP} Hierarchies for Polynomial Optimization on the Hypersphere},
  year   = {2012},
  eprint = {1210.5048},
  archivePrefix = {arXiv},
  url    = {https://arxiv.org/abs/1210.5048}
}

@article{FF21,
  author  = {Fang, Kun and Fawzi, Hamza},
  title   = {The Sum-of-Squares Hierarchy on the Sphere and Applications in Quantum Information Theory},
  journal = {Mathematical Programming},
  year    = {2021},
  volume  = {190},
  pages   = {331--360},
  doi     = {10.1007/s10107-020-01537-7},
  eprint  = {1908.05155},
  archivePrefix = {arXiv},
  primaryClass = {math.OC}
}

@article{LJ26,
  author  = {Lovitz, Benjamin and Johnston, Nathaniel},
  title   = {A Hierarchy of Eigencomputations for Polynomial Optimization on the Sphere},
  journal = {Mathematical Programming},
  year    = {2026},
  volume  = {219},
  pages   = {113--156},
  doi     = {10.1007/s10107-025-02251-y},
  eprint  = {2310.17827},
  archivePrefix = {arXiv},
  note    = {Published online July 21, 2025}
}

@inproceedings{HSS15,
  author    = {Hopkins, Samuel B. and Shi, Jonathan and Steurer, David},
  title     = {Tensor Principal Component Analysis via Sum-of-Square Proofs},
  booktitle = {Proceedings of the 28th Conference on Learning Theory},
  series    = {Proceedings of Machine Learning Research},
  volume    = {40},
  year      = {2015},
  pages     = {956--1006},
  publisher = {PMLR},
  url       = {https://proceedings.mlr.press/v40/Hopkins15.html}
}

@article{NS09,
  author  = {Nazarov, Fedor and Sodin, Mikhail},
  title   = {On the Number of Nodal Domains of Random Spherical Harmonics},
  journal = {American Journal of Mathematics},
  year    = {2009},
  volume  = {131},
  number  = {5},
  pages   = {1337--1357},
  doi     = {10.1353/ajm.0.0070},
  eprint  = {0706.2409},
  archivePrefix = {arXiv}
}

@article{LL15,
  author  = {Lerario, Antonio and Lundberg, Erik},
  title   = {Statistics on {Hilbert's} 16th Problem},
  journal = {International Mathematics Research Notices},
  year    = {2015},
  volume  = {2015},
  number  = {12},
  pages   = {4293--4321},
  doi     = {10.1093/imrn/rnu069},
  eprint  = {1212.3823},
  archivePrefix = {arXiv}
}
